\documentclass[useAMS,usenatbib]{mn2e}
\usepackage{graphicx}
\usepackage[usenames,dvipsnames]{color}
\usepackage{aas_macros}
\usepackage{lastpage}
\usepackage{comment}
\usepackage{subfigure}
\usepackage{hyperref}

\def\C06{\citet{Croton2006}}
\def\MCMC{Monte-Carlo Markov Chain}
\def\SAM{semi-analytic model}
\def\SMF{stellar mass function}
\def\BHBR{black hole--bulge relation}

\def\alphaSF{$\alpha_{\mathrm{SF}}$}
\def\epsilondisk{$\epsilon_{\mathrm{disk}}$}
\def\epsilonhalo{$\epsilon_{\mathrm{halo}}$}
\def\kappaAGN{$\kappa_{\mathrm{AGN}}$}
\def\fBH{$f_{\mathrm{BH}}$}
\def\gammaeject{$\gamma_{\mathrm{ej}}$}

\title[Constraining the evolution of semi-analytic models]{Constraining
the last 7 billion years of galaxy evolution in semi-analytic models}

\author[Mutch, Poole and Croton]{Simon J. Mutch$^{1,2}$\thanks{E-mail:
smutch@unimelb.edu.au}, Gregory B. Poole$^{1,2}$ and Darren J. Croton$^{1}$\\
$^{1}$Centre for Astrophysics \& Supercomputing, Swinburne University of
Technology, PO Box 218, Hawthorn, VIC 3122, Australia\\ 
$^{2}$School of Physics, University of Melbourne, Parkville, Victoria 3010,
Australia}

\begin{document}


\pagerange{\pageref{firstpage}--\pageref{lastpage}} \pubyear{2012}

\maketitle

\label{firstpage}

\begin{abstract} 
    We investigate the ability of the \C06{} semi-analytic model to reproduce
    the evolution of observed galaxies across the final 7 billion years of
    cosmic history.  Using \MCMC{} techniques we explore the available parameter
    space to produce a model which attempts to achieve a statistically accurate
    fit to the observed stellar mass function at $z{=}0$ and $z{\approx}0.8$, as
    well as the local black hole--bulge relation.  We find that in order to be
    successful we are required to push supernova feedback efficiencies to
    extreme limits which are, in some cases, unjustified by current
    observations.  This leads us to the conclusion that the current model may be
    incomplete.  Using the posterior probability distributions provided by our
    fitting, as well as the qualitative details of our produced stellar mass
    functions,  we suggest that any future model improvements must act to
    preferentially bolster star formation efficiency in the most massive halos
    at high redshift.
\end{abstract}


\section{Introduction}

Modern semi-analytic galaxy formation models are a commonly used tool to aid in
interpreting the statistical properties of large galaxy samples
\citep[e.g.][]{Kauffmann1999,Hatton2003,Croton2006,Bower2006,DeLucia2007,Somerville2008,Guo2011,Benson2012}.
In a $\Lambda$CDM universe, the physical properties of galaxies are largely
determined by the attributes of the halos in which they form, such as their mass
and merger history \citep{Mo1998}.  Semi-analytic models attempt to capture this
dependence, as well as the complex baryonic processes involved in galaxy
evolution, through a series of time evolving differential equations.  Free
parameters in the equations allow us to account for missing details in our
understanding and/or implementations of the relevant physics.

Traditionally, these parameters are `hand tuned' to accurately reproduce a small
subset of important observations, as well as achieve a reasonable level of
agreement with a larger number of other observed quantities.  In this way,
semi-analytic models have had considerable success in reproducing many of the
most basic statistical quantities of the local Universe such as the galactic
stellar mass function, the black hole--bulge relation, luminosity functions,
colour--stellar mass relations, Tully-Fisher relations and correlation
functions.

The procedure of manually calibrating model parameters can be extremely useful
in developing an intuition for the importance of each of the component physical
prescriptions and how they connect together.  However, it is often a challenging
and time-intensive task.  The quality of fit is usually assessed visually,
without providing a statistical measure of success.  Hence there is no way to
confirm that the chosen parameter values do truly provide the best possible
reproduction of the data, or indeed that they are unique.  Also, as the models
become more sophisticated the number of free parameters naturally grows, as does
the range of constraining observations.  These parameters can have complex and
highly degenerate interdependencies and, although the physically motivated
parametrisations give us a broad idea of what the major effects of each
parameter should be, it is extremely difficult to predict the exact consequences
of any changes on the full range of galaxy properties produced.  This problem is
ubiquitous in any flavour of galaxy formation simulation.

Fortunately, \SAM{}s are relatively computationally inexpensive, especially when
compared to full hydrodynamical galaxy formation simulations, and thus can be
run quickly.  This provides us with the ability to explore the parameter space
of these models in a sensible time frame, allowing us to not only find the
precise parameter values that produce the best match to the observable Universe,
but also understand the complex interplay between the included physical
processes.  As a result, there have been a number of attempts to automatically
calibrate \SAM{}s using Bayesian statistical tools such as Monte Carlo Markov
Chains \citep[MCMC; e.g.][]{Henriques2009,Lu2011,Lu2012}.

MCMC techniques have only relatively recently been applied to the task of
calibrating galaxy formation models, despite having been used extensively for a
number of years in other areas of astronomy such as cosmological parameter
estimation \citep[e.g.][]{Lewis2002}.   \citet{Kampakoglou2008} was the first,
constraining a fully analytic model of star formation against a number of
observations.  These included the cosmic star formation history and type-II
supernova rate out to high redshift.  They also applied a novel Bayesian
procedure to account for unknown systematics in their observational datasets.

In parallel to this work, \citet{Henriques2009} investigated the
\citet{DeLucia2007} semi-analytic galaxy formation model by calibrating it
against the redshift zero $K$-band luminosity function, colour--stellar mass
relation, and black hole--bulge relation.  Using their results, they were able
to draw conclusions about the interplay of the different parameters in their
model, as well as highlight some potential tensions in simultaneously matching
both the \BHBR{} and $K$-band luminosity function.  Following on from this,
\citet{Lu2012} calibrated a generic semi-analytic model \citep{Lu2011}, again
against the $z{=}0$ $K$-band luminosity function.  The large number of free
parameters and general construction of their model allowed them to mimic the
implementations of a number of different previously published models, and thus
to make more wide-reaching arguments about the success of semi-analytics in
general when attempting to replicate the observed Universe.  Using a method
outlined in \citet{Lu2011}, the authors also used the parameter probability
distribution to place uncertainties on a number of predictive quantities, both
in the local Universe and out to higher redshifts.

Rather than also implementing MCMC methods, \citet{Bower2010} introduced the
novel Bayesian technique of model emulation to calibrate the \citet{Bower2006}
semi-analytic model.  Their constraints were the $z{=}0$ $K$ and $b_J$-band
luminosity functions.  While model emulation provides a significantly better
scaling with large numbers of parameters than MCMC methods, the details of
its application are relatively complex.  In a companion paper, \citet{Benson2010}
implemented this technique to aid in calibrating a new and updated version of
the \citet{Bower2006} model against a large range of 21 different observations
and across multiple redshifts.  They also utilised 29 free parameters.  Their
aim, however, was not to provide an accurate statistical reproduction of each of
these observations, but to provide a final model which did a reasonable job of
qualitatively matching as many of them as possible.  Each of the observations
was therefore given an arbitrary weighting in the total likelihood calculation.
In addition, a last manual adjustment of the parameters was made to provide
their fiducial model.

In this paper, we use \MCMC{} techniques to statistically calibrate the
\citet{Croton2006} \SAM{}.  In particular we investigate whether the published
version of this model is capable of replicating not only the present day
Universe, but also the time evolution of the full galaxy population.  To achieve
this, we extend these previous works by considering the evolution of the
galactic stellar mass function between $z{=}0$ and $z{\approx}0.8$, but with
the restriction that we must simultaneously match the $z{=}0$ black hole bulge
relation.  Our work can most closely be compared with that of
\citet{Henriques2009}, as we use a similar model and one which is also run on
the merger trees constructed directly from N-body simulations.  However,
comparisons can also be made with the works of \citet{Lu2011,Lu2012} and
\citet{Bower2010}, given the similarities in the utilised modelling and analysis
techniques.

We emphasise that in all of the aforementioned studies to which our work can be
compared, the relevant models have only been constrained to match observations
of the local Universe.  While \citet{Lu2012} provides predictive quantities out
to high redshift to draw valuable conclusions about the validity of their model
across time, our work constitutes the first time that a robust statistical
calibration has been carried out at two redshifts simultaneously.  This allows
us to definitively test the ability of our model to reproduce the observed
growth of stellar mass in the Universe at $z{\la}1$.

This paper is laid out as follows:  In \S\ref{sec:method} we introduce
\MCMC{} methods and discuss some of the details of our particular
implementation.  In \S\ref{sec:SAM} we provide a brief overview of the
\citet{Croton2006} \SAM{}, focussing on the physical prescriptions that are of
particular relevance to this work.  We then move on to describing the
observational quantities we use to constrain our model in
\S\ref{sec:Obs_Constraints}.  Our results and analysis are presented in
\S\ref{sec:analysis}, with a detailed discussion of their significance
found in \S\ref{sec:discussion}.  Finally, we conclude by summarising our
main results in \S\ref{sec:conclusions}.

A standard $\Lambda$CDM cosmology with $\Omega_m{=}0.25$,
$\Omega_{\Lambda}{=}0.75$, $\Omega_{b}{=}0.045$ is utilised throughout this
work.  All results are quoted with a Hubble constant of $h{=}0.7$ (where
$h{\equiv}H_0/100\, \mathrm{kms^{-1}Mpc^{-1}}$) unless otherwise indicated.

\section{Method}
\label{sec:method}

In general, we wish to find the model parameter set with the highest statistical
likelihood as well as its uncertainty, given various observational constraints.
In theory, the most straightforward way of achieving this is to invoke the Law
of Large Numbers and draw independent samples from the joint posterior
probability distribution function (PDF) of the model parameters; this is the
probability of each parameter combination, given the set of constraining
observations.  Unfortunately, the presence of complex interdependencies between
the parameters means we often do not know the form of the complicated joint
posterior a-priori.

One way to overcome this problem is to implement \MCMC{} methods.  This
is a Bayesian statistical technique for probing complex, highly
degenerate probability distributions.  Specifically, we employ the commonly used
Metropolis--Hastings algorithm \citep{Hastings1970}.  In the following section,
we describe our particular implementation of this algorithm.  A more general
overview of MCMC and Bayesian techniques can be found in other works
\citep[e.g.][and references therein]{Lewis2002,Trotta2008}.

\subsection{MCMC implementation}
\label{sub:MCMCandSAMs}

Although far more efficient than simply probing the entire $N$--dimensional
parameter space with a regular grid, a MCMC chain still typically requires
many tens of thousands of propositions to fully sample the posterior.  This
necessitates short run times, on the order of a few seconds or less, for a
single realisation of the \SAM{}.  For this reason, we cannot run each model
iteration on the full dark matter merger trees of the entire input Millennium
simulation.  Instead we restrict ourselves to running on $1/512$ of the full
$0.125\,h^{-3}\mathrm{Gpc}^3$ volume.  This is equivalent to a comoving volume of
$2.44{\times}10^5\,h^{-3}\mathrm{Mpc}^3$.

Rather than choosing a contiguous sub-volume of the simulation to form our
input merger tree set, we randomly subsample an equivalent fraction of
the total number of merger trees.  This moderates the effects of cosmic
variance and also allows us to fully probe the halo mass function up to some
maximum limiting mass.  Note that we use the same merger tree sample for
every MCMC chain.  After making a number of technical changes to the code
base, we reduce the run time for a single input file from approximately 1.5
minutes to a just a few seconds with the \C06{} model running on 64 cores
of the Swinburne University of Technology's Green Machine\footnote{See
\url{http://www.astronomy.swin.edu.au/supercomputing/} for further details}.
These changes include load-balancing the input dark matter merger trees from
each individual file across multiple CPU cores, as well as removing costly
magnitude calculations.

For all of the results presented below, we combine two fully independent MCMC
chains, each with 100\,000 model calls in their integration phases.  This is
typically adequate to achieve well mixed and converged results except for
explicit cases which we specifically highlight in the text.  In order to assess
this we implement the Rubin--Gelman statistic \citep[assuming $\hat{R}{\la}1.03$
indicates convergence;][]{Gelman1992} as well as visually inspect the chain
traces.  We also run several other shorter test chains, all with different
random starting positions, in order to ensure that we are not missing any
discrete regions of high probability in our analysis.

\subsection{Principle Component Analysis}
\label{sub:PCA}

The primary product of an MCMC analysis is the posterior probability
distribution of the $N$-dimensional parameter space of the model, constrained
against the relevant observables.  This distribution contains a wealth of
information, not only about the highest likelihood parameter combination, but
also about the level to which each parameter is constrained and the degeneracies
that exist between them.  In order to aid with our interpretation of the
posterior distributions, we carry out a principal component analysis (PCA) when
appropriate.  This method compresses the information contained in the PDF into
as few basis vectors as possible.  In practice, the problem reduces to an
eigenvector decomposition of the covariance matrix, where the eigenvectors are
the principal components and the corresponding eigenvalues provide a measure of
the amount of variance they describe.  By carrying out such an analysis on a
MCMC chain, we are able to identify which parameters are responsible for
describing the bulk of the scatter in the posterior probability distributions.
Parameters which provide almost no variance in any of the principal components
can thus be interpreted as being well constrained by the relevant observations.

There are underlying assumptions and limitations associated with PCA\@ that
necessitate care in its interpretation.  This is particularly true in the case
of pathological PDFs exhibiting multiple discrete probability peaks or highly
non-linear degeneracies.  We must therefore be wary of placing strong emphasis
on the precise values obtained from such an analysis.  However, PCA does provide
a valuable tool for gaining a qualitative insight into which physical
prescriptions of the model are most important for matching particular
observations.  In some cases, a visual inspection of the PDF may indicate that a
parameter is well constrained by the relevant observations, however, a PCA
analysis could indicate that it is in fact small variations in the value of this
parameter that drives larger changes to the other parameters.  Also, if by
adding a new observational constraint the number of principal components
decreases, this indicates that the new constraint adds information that
successfully reduces the model freedom.

In order to carry out a principal component analysis of a posterior
distribution, the following steps are followed:  First, we take the integration
phase of the MCMC chain and calculate the mean value for each model parameter.
This value is then subtracted from all of the proposition vectors.  Next, a
covariance matrix is constructed and an eigenvector decomposition of this matrix
carried out.  Finally, the resulting eigenvectors are ranked in order of
decreasing eigenvalue.  As discussed above, the eigenvalues are a measure of the
amount of variance described by each eigenvector.  Deciding how many of the top
ranked eigenvectors form the principal component set is arbitrary; however, we
follow the common practice of including increasingly lower ranked vectors until
we have recovered 90\% or more of the total variance in our final set.

\section{The semi-analytical galaxy formation model}
\label{sec:SAM}

\def\Z{\vphantom{\parbox[c]{1cm}{\Huge DUMMY}}} 
\begin{table*}
\begin{minipage}{\textwidth}\centering
    \caption{The six free parameters of the semi-analytic model which we focus
        on in this work.  The original values of \C06{} are listed along with
        the best parameter values calibrated at $z{=}0$ (\SMF{} + \BHBR{}),
        $z{=}0$.83 (\SMF{}), and $z{=}0$ and 0.83 simultaneously.  The quoted
        uncertainties represent the 68\% confidence limits of the marginalised
        posterior distributions where appropriate.  See \S\ref{sec:SAM} for a
        description of the role played by each parameter.}
  \label{tab:params}%
  \begin{tabular}{|c|c|c|c|c|c|}
    \hline
    Parameter name & Physical prescription & Original value &
    \multicolumn{3}{|c|}{Best parameter values} \\
    &&& $z=0$ & $z=0.83$ & $z=0+0.83$\\ 
    \hline\hline
    
\Z  $\alpha_{SF}$ & In-situ star formation & 0.07 &
    $0.019_{-0.003}^{+0.003}$ & $0.044_{-0.019}^{+0.033}$ &
    $0.055_{-0.016}^{+0.011}$\\ 
    
\Z  $\epsilon_{disk}$ & SN feedback &
    3.5 & $5.14_{-0.76}^{+1.22}$ & $4.79_{-0.94}^{+1.87}$ & 
    $13.8_{-2.2}^{+4.1}$\\ 
    
\Z  $\epsilon_{halo}$ & SN feedback &
    0.35 & $0.26_{-0.03}^{+0.06}$ & $0.41_{-0.06}^{+0.16}$ &
    $1.18_{-0.20}^{+0.38}$\\ 
    
\Z  $\gamma_{ej}$ & Gas Reincorporation & 0.5 &
    $7.1_{-4.8}^{+4.9}\times 10^{-3}$ & $7.1_{-4.2}^{+1.0}\times 10^{-3}$ &
    $1.13_{-0.24}^{+0.30}$\\ 
    
\Z  $f_{BH}$ & Black hole growth & 0.03
    & $0.015_{-0.003}^{+0.002}$ & $0.015_{-0.010}^{+0.035}$ & 
    $0.025_{-0.007}^{+0.007}$\\ 
    
\Z  $\kappa_{AGN}$ & Black hole growth &
    $5.89\times10^{-4}$ & $1.90_{-0.33}^{+0.39}\times10^{-4}$ &
    $1.71_{-1.21}^{+4.13}\times 10^{-4}$ &
    $1.47_{-0.67}^{+0.24} \times 10^{-4}$\\ 
    
    \hline
  \end{tabular}
\end{minipage}
\end{table*}

In this section we describe the \C06{} \SAM{} used in this work.  This
model has a number of free parameters which regulate a broad range
of physical processes from black-hole accretion and feedback, to the effects
of cosmic re-ionisation.  However, as in \citet{Henriques2009}, we focus only on
the six main parameters which regulate star formation, super-nova feedback
and black hole growth (Table~\ref{tab:params}).  These are less well constrained
by observation or theory than many of the other model parameters
\citep[c.f.][Table 1]{Croton2006}, and are strongly dependant on the particular
implementations of the physical processes.

The remainder of this section is devoted to outlining the role that each of
these six free parameters play in shaping the properties of the galaxy
population.  For a more detailed description that includes all of the physical
prescriptions present in the model, the reader is referred to \C06{}.  Those
already familiar with the model can forgo this section.

\subsection{Star formation and supernova feedback}
\label{sub:SFandSN}

In accordance with the work of \citet{Kennicutt1998}, star formation is
regulated by a critical surface density of cold gas.  This is in turn related to
the radius of the galaxy disk using the empirical relation of
\citet{Kauffmann1996}.  Whenever the mass of cold gas ($m_{\mathrm{cold}}$)
exceeds the critical mass suggested by this relation ($m_{\mathrm{crit}}$), a
burst of star formation occurs.  The star formation rate ($\dot{m}_{*}$) is then
given by:
\begin{equation}
\dot{m}_{*} = \alpha_{SF} (m_{\mathrm{cold}} - m_{\mathrm{crit}}) /
t_{\mathrm{dyn,disk}}\ ,
\end{equation}
where $t_{\mathrm{dyn,disk}}$ is the dynamical time of the disk and \alphaSF{}
is a free parameter controlling the efficiency at which the excess cold gas is
converted into stars over this timescale.

With each new star formation episode, the highest mass stars will rapidly
evolve and end their lives as energetic supernovae.  The injection of this
energy into the galaxy interstellar medium will heat up a fraction of the cold
gas, expelling it from the plane of the disk and into the surrounding hot
halo component.  The amount of cold gas reheated in this way follows:
\begin{equation}
\Delta m_{\mathrm{reheated}} = \epsilon_{\mathrm{disk}} \Delta m_{*}\:.
\end{equation}

The parameter \epsilondisk{} is equivalent to the supernova wind mass loading
factor with \citet{Croton2006} fixing its value to be 3.5 based on the
observations of \citet{Martin1999}.

The amount of energy released per unit mass over the relevant time interval is
approximated by:
\begin{equation}
    \Delta E_{\mathrm{SN}} = 0.5 \epsilon_{\mathrm{halo}} V^2_{\mathrm{SN}}
    \Delta m_{*}\ ,
\end{equation}
where $0.5 V^2_{\mathrm{SN}}$ is the mean energy injected by supernova per unit
mass of star formation and the parameter \epsilonhalo{} controls the efficiency
with which this energy can actually reheat the disk gas.

The amount of energy required to adiabatically reheat
$\Delta m_{\mathrm{reheated}}$ of cold gas and add it to the hot halo reservoir
is given by:
\begin{equation}
   \Delta E_{\mathrm{hot}} = 0.5\Delta m_{\mathrm{reheated}}V_{\mathrm{vir}}^2\
   ,
\end{equation}
where $V_{\mathrm{vir}}^2$ is the virial velocity of the host dark matter halo
and $0.5 V_{\mathrm{vir}}^2$ is the thermal energy per unit mass of the hot halo
component.  If $\Delta E_{\mathrm{excess}}{=} \Delta E_{\mathrm{SN}} {-} \Delta
E_{\mathrm{reheated}}$ is positive then enough energy is provided to physically
eject some fraction of the mass from the system entirely:
\begin{equation}
    \label{eqn:meject}
    \Delta m_{\mathrm{ejected}} = \frac{\Delta
	E_{\mathrm{excess}}}{E_{\mathrm{hot}}} m_{\mathrm{hot}} = \left(
	\epsilon_{\mathrm{halo}} \frac{V^2_{\mathrm{SN}}}{V^2_{\mathrm{vir}}} -
	\epsilon_{\mathrm{disk}} \right) \Delta m_{*}\:.
\end{equation}

This ejected gas is added to an external reservoir of material from where it
plays no further role in the current heating/cooling cycle. As the dark matter
halo grows, some of this ejected material may fall back into the deepening
potential well and will be added back into the hot halo component.  The fraction
of ejected material that is re-incorporated per halo dynamical time is
controlled by the parameter \gammaeject:
\begin{equation}
    \label{eqn:reincorporation}
    \dot m_{\mathrm{ejected}} = -\gamma_{\mathrm{ej}}
    m_{\mathrm{ejected}}/t_{\mathrm{dynamical}}\:.
\end{equation}

\subsection{Supermassive black hole growth and feedback}
\label{sub:SMBHs}

As discussed by \cite{Croton2006}, eqn.~\ref{eqn:meject} implies that for
galaxies in halos with $V_{\mathrm{vir}} {>}
\epsilon_{\mathrm{halo}}/\epsilon_{\mathrm{disk}}\,V^2_{\mathrm{SN}}$, supernova
feedback processes are unable to eject any material from the galaxy--halo
system.  For their choice of parameters, this corresponds to dark matter halos
with $V_{\mathrm{vir}}{\ga} 200\, \mathrm{km\,s^{-1}}$.  In systems more massive
than this supernova feedback becomes inefficient at suppressing the long term
cooling of gas and associated star formation.  The result is an over prediction
of the number of high mass galaxies in the Universe.  The inclusion of feedback
effects from super massive central black holes provides a well motivated and
physically plausible mechanism for further regulating the cooling of gas in
these high mass systems.

Central black holes grow via two mechanisms in our model.  The first is the
`quasar' mode which results from galaxy merger events.  During such an event,
the progenitor black holes are assumed to coalesce with no loss of
mass due to dissipative processes.  A fraction of the cold gas of the progenitor
galaxies is also accreted by the newly formed central black hole, increasing
its mass further:

\begin{equation}
    \label{eqn:BHquasarmode}
    \Delta m_{\mathrm{BH,quasar}} = \frac{f_{\mathrm{BH}}\,m_{\mathrm{cold}}
    \,m_{\mathrm{sat}} / m_{\mathrm{central}}} {1
	+ (280 \mathrm{km\,s^{-1}}/V_{\mathrm{vir}})^2}\ ,
\end{equation}
where \fBH{} is a free parameter.  This is the dominant growth mechanism for
black holes in our model, although it is important to note that this growth is
not accompanied by the injection of energy into the inter-stellar medium.

Black holes are also allowed to grow quiescently through the continual accretion
of hot gas in what is called the `radio' mode.  This is characterised by the
following simple model:
\begin{equation}
    \label{eqn:BHradiomode}
    \dot m_{\mathrm{BH,radio}} = \kappa_{\mathrm{AGN}} \left(
    \frac{m_{\mathrm{BH}}}{10^8 \mathrm{M_{\sun}}} \right) \left(
    \frac{f_{\mathrm{hot}}}{0.1} \right) \left(
    \frac{V_{\mathrm{vir}}}{200\:\mathrm{km\,s^{-1}}} \right)^3\ ,
\end{equation}
where \kappaAGN{} is our last free parameter and $f_{\mathrm{hot}}$ is the
fraction of the dark matter halo mass in the hot component.  In contrast to the
quasar mode, here material accreted by the black hole results in the injection
of energy directly into the inter-stellar medium:
\begin{equation}
    \label{eqn:BHfeedback}
    L_{\mathrm{BH}} = \eta \dot m_{\mathrm{BH}}c^2\:.
\end{equation}
The effect is a reduction, or even complete cessation, of cooling onto the disk.
By regulating the availability of cold gas in massive galaxies, this feedback
mechanism is able to efficiently reduce the normalisation of the massive end of
the stellar mass function \citep[c.f.][figure 8]{Croton2006}.

\section{Observational Constraints}
\label{sec:Obs_Constraints}

In this section we provide an overview of the observational constraints used in
our analysis: the stellar mass function and black hole-bulge relation.  We also
discuss the statistical tests we employ to asses the quality of the
reproduction achieved by our model.  We implement the \SMF{} constraint at both
$z{=}0$ and $z{=}0.83$, and the \BHBR{} constraint at $z{=}0$ alone.  This
allows us to test if our model can not only reproduce the local and high
redshift universes independently, but also if it can be successful at
constraining the late time evolution of the galaxy population between these two
epochs.

\subsection{The Stellar Mass Function}
\label{sub:SMF_constraint}

The \SMF{} is a fundamental observable in the study of galaxy formation and
evolution.  It provides one of the most basic statistical descriptions of the
galaxy population -- the number of galaxies per unit stellar mass, per unit
volume ($\phi$) as a function of stellar mass ($M_*$) -- and is directly
influenced by the full range of physical processes associated with the evolution
of the galaxy population.  It is therefore important for any successful galaxy
formation model to be able to provide a realistic reproduction of this quantity.

For both our low and high redshift \SMF{}s, we have invested a great deal of
effort to use the most suitable observations that permit the use of accurate
uncertainties in our analysis.  In order to fairly judge the ability of a model
to reproduce any observational constraint, it is extremely important that the
observations have realistic uncertainties.  If these are under estimated, then
the model likelihood will be unfairly punished for predicting slight deviations;
if they are overestimated then the constraints on the model parameters will be
poor. 

\subsubsection{Low redshift}
\label{sub:lowzSMF}

There are a large number of local measurements of the galaxy \SMF{} available in
the literature \citep[e.g.][]{Cole2001,Baldry2004,Panter2007}.  Typically,
stellar masses are inferred in these works through the use of empirically
determined stellar mass--light ratios.  Unfortunately, masses estimated in this
way require the use of a number of implicit assumptions regarding the stellar
initial mass function (IMF), star formation histories, and the integrated
effects of dust extinction.  As a result, these masses can often suffer from
large systematic uncertainties \citep{Conroy2009} which can be difficult to
quantify and are often not included in published \SMF{}s.

For this work, we utilize the $z{=}0$ \SMF{} of \citet{Baldry2008}.  The main
advantage of this particular work, for our purposes, is that the quoted
uncertainties include an estimate of the systematic contributions associated
with the use of colour dependant mass-light ratios, as discussed above, as well
as the usually considered Poisson uncertainties.  This was achieved by
considering the mass function produced using a range of different stellar mass
determinations, aggregated from five independent works, of
matching galaxies drawn from the Sloan Digital Sky Survey
\citet[SDSS][]{York2000} New York University Value-Added Galaxy Catalogue
\citep[NYU-VAGC][]{Blanton2005}.

In order to directly compare our model to these observations, we convert the
averaged \citet{Kroupa2001} and \citet{Chabrier2003} IMF used by
\citet{Baldry2008} to the \citet{Salpeter1955} IMF assumed by our model.  This
is done by applying a systematic shift of +0.26 dex to the stellar mass values
of the observed \SMF{}\@.

The particle mass of the Millennium simulation, from which our input dark matter
merger trees are generated, is $8.6{\times}10^8\,\mathrm{M_{\sun}}h^{-1}$.
Typically, ${\sim} 100$ particles are required to attain well resolved,
non-stochastic merger histories for a dark matter halo.  Using the default
published model of \C06{}, this corresponds to galaxies with stellar masses of
$\log_{10}(h^{-2}M/\mathrm{M_{\sun}}){\approx} 9.5$.  Since we are using only
$1/512$ of the full simulation volume in our analysis, we are unable to fully
average out the stochastic nature of the properties of galaxies below this mass
and thus we use this as a conservative lower limit on the reliability of stellar
masses generated by the model.  We reflect this in our analysis by cutting our
constraining observations to only include stellar masses above this lower limit.

\begin{figure}
    \includegraphics[width=\columnwidth]{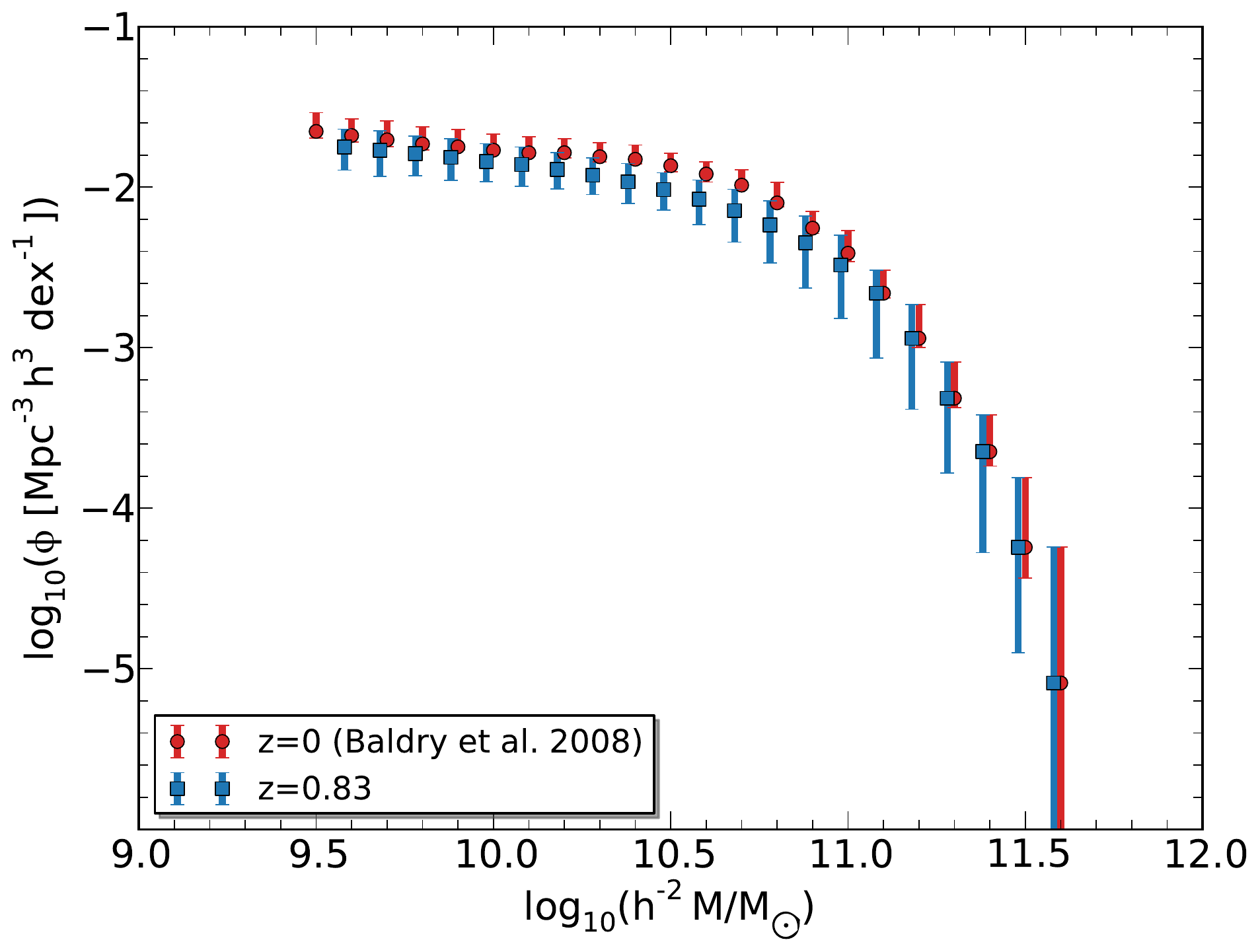}
    \caption{\label{fig:z0.83_SMF_constraint} The observed $z{=}0$ \citep[red
      circles;][]{Baldry2008} and $z{=}0$.83 (blue squares) stellar mass
      functions with 68\% confidence limits employed as model constraints in
      this work.  For clarity, the $z{=}0.83$ data has been shifted by $-0.02$
      dex in stellar mass.  The $z{=}0$.83 values and associated uncertainties
  are a result of homogenising and combining several published Schechter
functions (see \S\ref{sub:SMF_constraint} for details).}
\end{figure}

\subsubsection{High redshift}
\label{sub:highzSMF}

In this paper we also constrain the model using MCMC at redshifts greater than
zero.  Unfortunately, it is extremely challenging to measure the observed
stellar mass function at high redshift.  To fully sample both the low and high
mass tails of the distribution simultaneously one requires a survey sample of
both high depth and large volume.  In addition to this, the systematic
uncertainties associated with assumptions such as star formation histories
become even larger at increasingly higher look-back times, and again, these
systematics are often excluded from any quantitative analysis in the literature.
It is therefore unsurprising that many published $z{\ga}0.5$ \SMF{}s
display significant disagreement, sometimes even at the level of 2$\sigma$.

In order to obtain a single $z{\approx}0.8$ \SMF{} which provides a reasonable
estimate of the systematics, we create a weighted average from a number of
recently published Schechter function fits in the literature:
\citet{Drory2009}($z{=}0.8$--1.0), \citet{Ilbert2010}($z{=}0$.8--1.0),
\citet{Pozzetti2007}($z{=}0.7$--9.0).  Each was converted from a
\citet{Chabrier2003} IMF to a \citet{Salpeter1955} IMF using a constant offset
of $+0.24\, \log_{10}(M_*/\mathrm{M_{\sun}})$ in stellar mass.  All values were
also homogenised to $h$=1 and the mass functions cut at the relevant mass
completeness limits of each sample.  \citet{Pozzetti2007} provides four mass
function fits calculated from the VVDS \citep[Visible Imaging Multi-Object
Spectrograph;][]{Le-Fevre2005} survey using two different sample selection
criteria and two alternative star formation history models.  In total, we
therefore employ six observed \SMF{}s covering a redshift range of
$z{=}0.7$--1.0.

Our final result was calculated by averaging these homogenised observations to
provide a single mass function at a mean redshift of $z{=}0.83$.  This was
done as follows:  All of the utilised mass functions provide $\pm\, 1\sigma$
uncertainties on the best fitting Schechter function parameter values.  We
incorporate these by taking each Schechter function in turn and generating 1000
realisations with parameter values randomly sampled from appropriate probability
distributions.  Ideally these distributions would account for the covariance
which exists between the different fitted parameters, but this information was
not provided in the relevant publications.  Instead we sampled
Gaussian (or skewed Gaussian) distributions centered on the best fit values and
with their quoted standard deviations.  The mean and $1\sigma$ uncertainties of
$\phi$ in each stellar mass bin are then calculated using the random
realisations from all six of the observed input functions.  To ensure
consistency with the $z{=}0$ \SMF{} of \citet{Baldry2008} we demand that
$\phi$ and its upper uncertainty is less than or equal to the respective
$z{=}0$ values at all stellar masses.  Such a restriction is well justified
given that the observed total stellar mass density is found to decrease by
approximately a factor of a half between $z{=}0$--1 \citep{Drory2009}.

Our final aggregated $z{=}0.83$ \SMF{} is shown in
Fig.~\ref{fig:z0.83_SMF_constraint}.  As a simple first order check of its
validity, we confirm that the integrated stellar mass density, over the range of
stellar masses present, is $0.64^{+0.21}_{-0.19}$ times that of the $z{=}0$ value.
The upper and lower bounds here account for the uncertainty in the mass
functions at both redshifts.  This shows broad agreement with observational
results \citep[e.g.][]{Marchesini2009}.  For comparison,
Fig.~\ref{fig:z0.83_SMF_constraint} also displays the constraining $z{=}0$
observations of \citet{Baldry2008}.

In order to calculate the likelihood of the model \SMF{}s, given the
observational data, we use a simple $\chi^2$ statistic.  For a single
stellar mass bin:
\begin{equation}
    \label{eqn:L_SMF}
    \mathcal{L}_{\mathrm{SMF}}(\theta) \propto \exp(-0.5\chi^2(\theta)) = \exp\left(
    -\frac{1}{2}\frac{(\phi_{\mathrm{obs}} - \phi_{\mathrm{mod}}(\theta))^2}
    {\sigma_{\mathrm{obs}}^2+\sigma^2_{\mathrm{mod}}(\theta)}\right)\:,
\end{equation}
where $\theta$ is the set of model parameters used and $\sigma$ represents the
associated uncertainties in each measurement.  We estimate
$\sigma_{\mathrm{mod,i}}$ using Poisson statistics to be
$\sqrt{n_i}\,h^3\mathrm{Mpc^{-3}dex^{-1}}$, where $n_i$ is the number of model
galaxies in bin $i$.  

We note that a number of previous works which have calibrated \SAM{}s using MCMC
techniques have tended to favor the use of the $K$--band luminosity function as
their primary constraint instead of the \SMF{} \citep{Henriques2009,Lu2012}.
As the $K$--band is well known to be a good proxy for stellar mass, both quantities
provide comparable constraints on the galaxy population.  As discussed above, it
can be difficult to derive accurate stellar masses for observed galaxies due to
the degeneracies and poorly understood systematics of dust attenuation,
mass--light ratios and IMFs.  Luminosity functions are, however, directly
observable and it is for this reason that they have been adopted by previous
works.  Unfortunately, producing a luminosity function from a \SAM{} involves
many of the same poorly understood physics and systematic uncertainties.
Specifically, we must include assumptions about dust attenuation and stellar
population synthesis models, in order to convert model stellar masses to
luminosities. 

As discussed previously, having realistic estimates of the relevant
uncertainties is important for our MCMC analysis.  Thus, we prefer to implement
the \SMF{} as the primary constraint, due to the availability of a number of
works which provide a quantitative analysis of some of the uncertainties
associated with measuring a stellar mass function at various redshifts
\citep[e.g.][]{Baldry2008,Pozzetti2007,Marchesini2009}.  Although it is true
that these uncertainties may still be underestimated, a similar estimate of the
systematics associated with a model derived luminosity function is beyond the
scope of this paper.

\subsection{The Black Hole--Bulge Relation}
\label{sub:BHBR_constraint}

It is well established that the masses of central super-massive black holes show
direct correlations with the properties of their hosts' bulges
\citep[e.g.][]{Magorrian1998,Haring2004,Sani2011}.  This suggests a physical
connection between the mass growth of these two components.  Given the
importance of AGN feedback in shaping the galaxy population, especially for high
galaxy masses at $z{<}1$, it is important that our model be able to reproduce this
observed relation.  This is especially so if we wish to use the model to make
any predictions for how black holes of different masses populate different
galaxy types.

Similarly to \citet{Henriques2009}, we implement the observations of
\citet{Haring2004} as our constraint for the $z{=}0$ \BHBR{}.  Their sample is
comprised of 30 nearby galaxies (the majority of objects being ${\la} 42\,
h^{-1}\mathrm{Mpc}$ away) with the bulge and black hole masses sourced from a
number of different works.

Observationally, it is still unclear whether or not there is a significant
evolution in the black hole--bulge relation between $z{=}0$ and $z{=}1$.  In
general, an evolution is predicted by theory \citep[e.g.][]{Croton2006b}, and is
tentatively measured by a number of authors
\citep[e.g.][]{McLure2006,Merloni2010}.  Unfortunately, observations of $z{>}0$
\BHBR{}s are generally hampered by systematic uncertainties which, when
included, make the significance of a deviation from the null hypothesis of no
evolution much less certain \citep{Schulze2011}.  For this reason, we choose not
to implement a \BHBR{} constraint at $z{=}0.83$.

To asses the likelihood of our model fit to the data, we implement the same
likelihood calculation as that of \citet{Henriques2009}.  First, the galaxy
sample is segregated into two bins defined by lines perpendicular to the best
fit relation of \citet{Haring2004}: 
\begin{equation}
\log_{10}(M_{\rm BH}) =
-0.89(\log_{10}(M_{\rm bulge}/{\rm M_{\sun}})-11) +{\rm offset}
\end{equation}
where ${\rm offset{=}[5.39,8.2,12.23]}$.  The binomial probability theorem is then
used to calculate what the likelihood is of finding the ratio of
\textit{observed} galaxies above and below the \citet{Haring2004} best fit line
in each bin:
\begin{equation}
    \label{eqn:L_BHBR}
    \mathcal{L}_{\mathrm{BHBR}} = \left\{
        \begin{array}{l l}
            2I_p(k, n-k+1) & \quad Ip\le0.5\\
            2(1-I_p(k, n-k+1)) & \quad Ip>0.5\\
        \end{array} \right.\:,
\end{equation}
where $k$ is the number of \textit{observed} galaxies above the best fit line in
each bin, $n$ is the total number of galaxies in the bin, and $p(\theta)$ is the
fraction of galaxies above the best fit line from the \textit{model}.
$I_p(x,y)$ is the regularised incomplete gamma function.  As described in
\citet{Henriques2009}, the reason for using two formulae with conditions is to
ensure that any excess of galaxies both above {\it and}\ below the best fit line
results in a low likelihood (i.e.\ both tails of the distribution).

\section{Analysis}
\label{sec:analysis}

In this section we present our main analysis.  First, we investigate the
restrictions placed on the model parameters by the individual observations at
$z{=}0$ and $z{=}0.83$.  This allows us to test which parameters are most
strongly constrained by each observation as well as identify any tensions
between these constraints.  The findings are then used to guide our
interpretation when calibrating the model against all three of our constraints
at both redshifts simultaneously (\S\ref{sub:multi_z_constraints}).

\subsection{Redshift zero}
\label{sub:z0_analysis}

\subsubsection{The Stellar Mass Function}
\label{sub:z0_smf_analysis}

\begin{figure*}
    \begin{minipage}{0.99\textwidth}
    \includegraphics[width=\columnwidth]{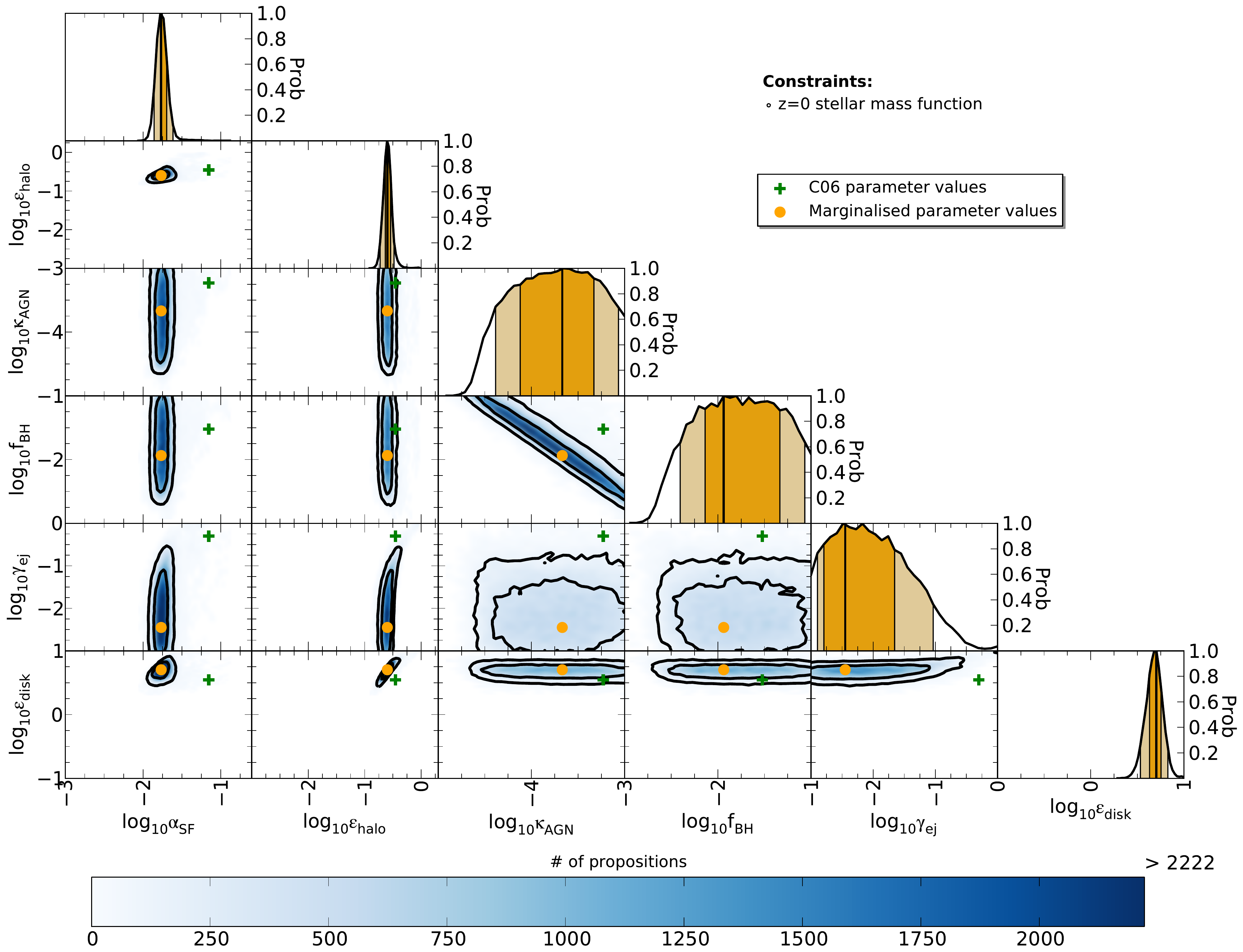}
    \caption{\label{fig:z0_smf_prob_matrix} 
    2-D posterior probability distributions (off-diagonal panels) for all
    combinations of the six free model parameters when constraining the model
    against the {\it $z{=}0$ \SMF{} alone}.  Black lines indicate the 1 and 2-$\sigma$
    confidence contours.  The limits of each panel indicate the prior ranges.
    Orange circles show the location of the marginalised best values of
    each parameter.  The histograms in the diagonal panels represent the 1-D
    marginalised probability distributions, with the 1 and 2$\sigma$ confidence
    intervals shown by the dark and light shaded regions respectively.}
\end{minipage}
\end{figure*}

We begin by considering the $z{=}0$ \SMF{} and investigate the restrictions
placed on the model parameters by this constraint alone.  The histograms on the
diagonal panels of Fig.~\ref{fig:z0_smf_prob_matrix} display the 1-dimensional
marginalised posterior distributions for each of the six free parameters.  The
highly peaked, Gaussian-like distributions of the star formation (\alphaSF{}),
supernova halo gas ejection (\epsilonhalo{}), and supernova cold gas reheating
(\epsilondisk{}) efficiencies indicate that these are well constrained by the
observed $z{=}0$ \SMF{} alone.  Conversely, the wide and relatively flat
distributions of the merger driven black hole growth (\fBH{}) and radio mode AGN
heating (\kappaAGN{}) efficiencies, suggest that their precise values are not
particularly well constrained.  The remaining off-diagonal panels of
Fig.~\ref{fig:z0_smf_prob_matrix} show the 2-dimensional posterior probability
distributions for all combinations of the six free model parameters (blue shaded
regions).  The black contours indicate the associated 1 and 2-$\sigma$
confidence intervals.

Although the $z{=}0$ \SMF{} alone does not allow us to say what the precise
values of the merger driven black hole growth efficiency (\fBH{}) and radio mode
AGN heating efficiency (\kappaAGN{}) must be, it does place a strong
constraint on their ratio.  This is indicated by the 2-D posterior distribution
of \fBH{} vs. \kappaAGN{} which shows a strong correlation between the allowed
values of these two parameters.  This is a direct consequence of the degeneracy
between central black hole mass (which is dominated by quasar mode growth and
thus \fBH{}) and the value of \kappaAGN{} in determining the level of radio mode
heating (c.f.~Eqn.~\ref{eqn:BHradiomode}).  This heating plays a key role in
shaping the high mass end of the stellar mass function where supernova feedback
becomes ineffective at regulating the availability of cold gas.  A similar
degeneracy was also noted by \citet{Henriques2009} when constraining their model
against the observed K-band luminosity function.

A principal component analysis of the joint posterior suggests that its variance
can be understood predominantly through the combination of two equally weighted
principal components.  The star formation efficiency (\alphaSF{}) and supernova
halo gas ejection efficiency (\epsilonhalo{}) provide almost no contribution to
the variance in either component, indicating that both are truly well
constrained by the \SMF{}.  On the other hand, the value of the ejected gas
reincorporation rate parameter (\gammaeject{}) does contribute significantly to
both components.  Interestingly, the supernova cold gas reheating efficiency
(\epsilondisk{}) also makes a dominant contribution to one of the principal
components, suggesting that, although it appears well constrained in
Fig.~\ref{fig:z0_smf_prob_matrix}, small variations about the mean can be
accommodated by a combination of changes to the remaining parameters controlling
black hole growth and AGN radio mode feedback (\fBH{} and \kappaAGN{}).

As well as investigating the parameter constraints and degeneracies, we also
wish to know what single set of parameters provides us with the best overall
reproduction of the relevant observations.  The orange points in
Fig.~\ref{fig:z0_smf_prob_matrix} indicate the marginalised best parameter
values.  This is the parameter set around which there was the largest number of
accepted propositions in the MCMC chain.  These values, along
with their 68\% confidence limits, are presented in Table~\ref{tab:params}.  

\begin{figure}
    \includegraphics[width=\columnwidth]{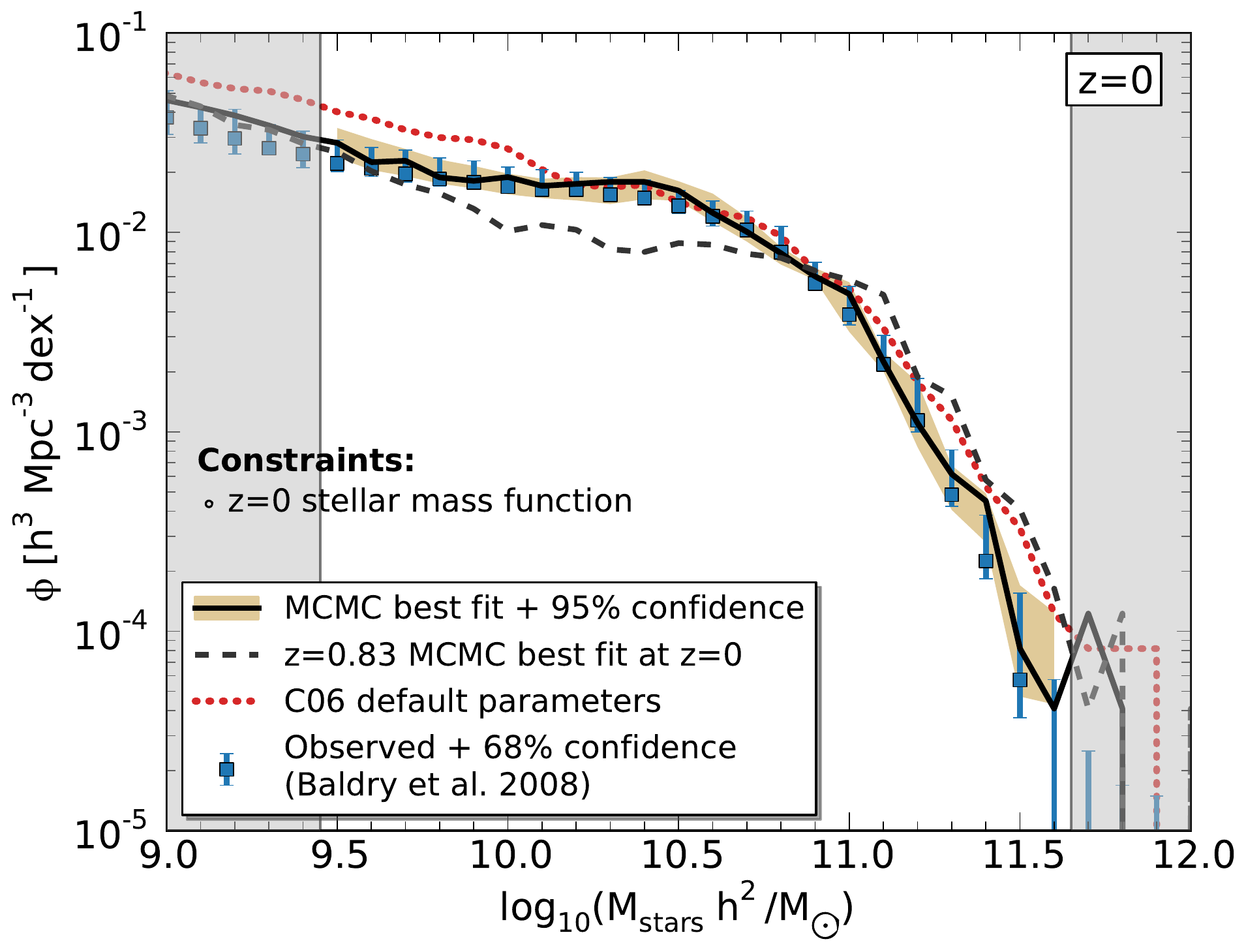}
    \caption{\label{fig:z0_smf} 
      The $z{=}0$ \SMF{} resulting from the best fit parameter values, as
      determined by constraining the model against the observed {\it $z{=}0$ \SMF{}
    alone}.  The solid line with shaded region shows the model result, along with
    the 95\% confidence limits calculated from the posterior distribution.  Blue
    error bars indicate the constraining observations and 68\% confidence regions of
    \citet{Baldry2008}.  The default \C06 prediction is shown by the red dotted
    line.  Only stellar masses in the unshaded region of the plot were used to
    constrain the model. The model prediction when using the best fit parameters
    constrained against the $z{=}0$.83 \SMF{} is also shown for comparison (black
    dashed line; \S\ref{sub:highz_analysis}).}
\end{figure}

In Fig.~\ref{fig:z0_smf} we show the \SMF{} produced by the model using these
best fit parameters, as well as the constraining observations of
\citet{Baldry2008} and the model prediction using the default \C06 parameters.
The orange shaded region encompassing the best fit line indicates the associated
95\% confidence limits.  These are calculated using all of the mass functions
produced during the integration phase of the MCMC chain.  When compared to the
original \C06 results, the best fit model more accurately reproduces the
distribution over the full range of masses - in particular the dip and
subsequent rise in galaxy counts that occurs around
$10^{10}\:h^2\mathrm{M^{-1}_{\sun}}$.  

\subsubsection{The Black Hole--Bulge Relation}
\label{sub:z0_BHBR_analysis}

\begin{figure}
    \includegraphics[width=\columnwidth]{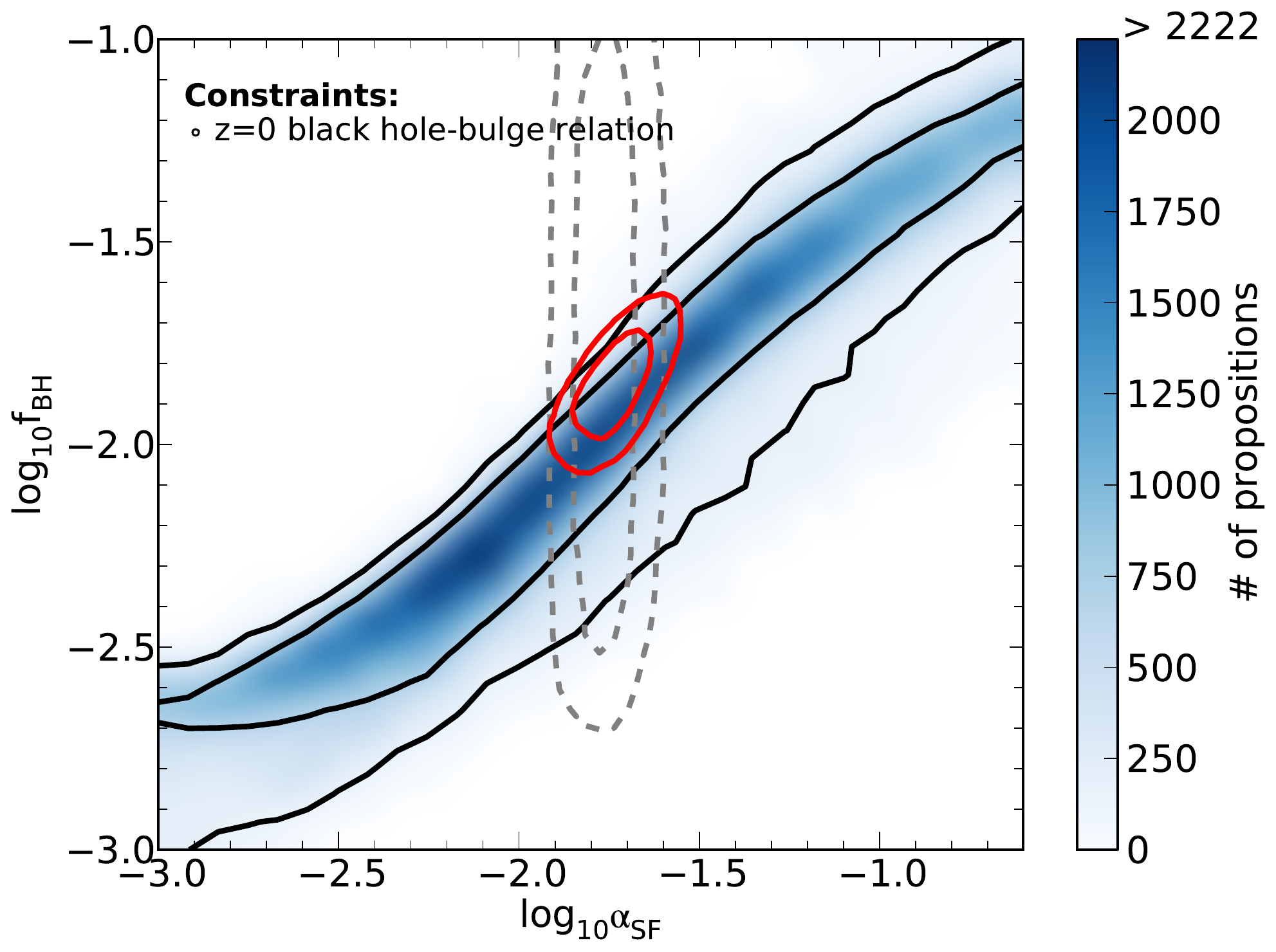}
    \caption{\label{fig:bhbr_prob_panel} The \fBH{}--\alphaSF{} posterior
    distribution (blue shaded) when constraining the model against the {\it
    $z{=}0$ \BHBR{} alone}.  Black lines represent the 1 and 2$\sigma$
    confidence contours.  Grey dashed lines show the equivalent confidence
    contours found when constraining the model against the $z{=}0$ \SMF{}
    alone (Fig.~\ref{fig:z0_smf_prob_matrix}), whilst red solid lines show the
    contours found when constraining against the $z{=}0$ \SMF{} and \BHBR{}
    simultaneously.}
\end{figure}

In order to break the above degeneracy between the merger driven black hole
growth and radio mode AGN heating efficiencies (\fBH{} and \kappaAGN{}), we
require the addition of another constraint which directly ties the properties of
the central black holes to those of the galaxies in which they form.  Following
\citet{Henriques2009}, we turn to the observed black hole--bulge mass relation
for this purpose.  Unlike the \SMF{}, which provides strong constraints in a
number of parameter planes, the \BHBR{} only constrains the \fBH{}--\alphaSF{}
(star formation efficiency) plane.

The utility of this particular constraint can be traced to the fact that
it provides a relation between the mass of the central black hole and
spheroidal component of a galaxy.  Bulges can grow in the model via two
different mechanisms.  The first is through merger events.  However, none of the
six free model parameters directly influence the strength of this mechanism.
The second method of bulge growth is via disk instabilities.  We treat this
using a modified version of the simple, physically motivated prescription of
\citet{Mo1998} whereby, once the surface density of stellar mass in the disk of
a galaxy becomes too great, the disk becomes unstable.  In this situation, a
fraction of the disk stellar mass is transferred to the bulge component in order
to restore stability. Hence, bulge growth via this mechanism is regulated by the
amount of stars already present in the disk as well as the mass of new stars
forming at any given time.  These, in turn, are modulated by the efficiency of
star formation (\alphaSF{}).  Black holes, on the other hand, gain the majority
of their mass via the merger driven quasar mode which is regulated by \fBH{}.

The posterior probability distribution for the \BHBR{} constraint alone is shown
in Fig.~\ref{fig:bhbr_prob_panel}.  As expected, increasing the efficiency of
star formation (\alphaSF{}), and therefore the growth of bulges through disk
instabilities, requires an increase in the efficiency of black hole growth
(\fBH{}).  Although omitted here for brevity, the constraints provided by this
observation on the three parameters which modulate star formation and supernova
feedback, are extremely weak.  However, in the case of the supernova halo gas
ejection efficiency parameter (\epsilonhalo{}), the marginalised posterior
distribution only overlaps with those of the stellar mass function constraint to
within $2\sigma$.  In other words, there is a slight tension between the
parameter sets favoured by the \BHBR{} and \SMF{}.

\begin{figure*}
    \begin{center}
        \begin{minipage}{0.99\textwidth}
	    \includegraphics[width=0.99\textwidth]{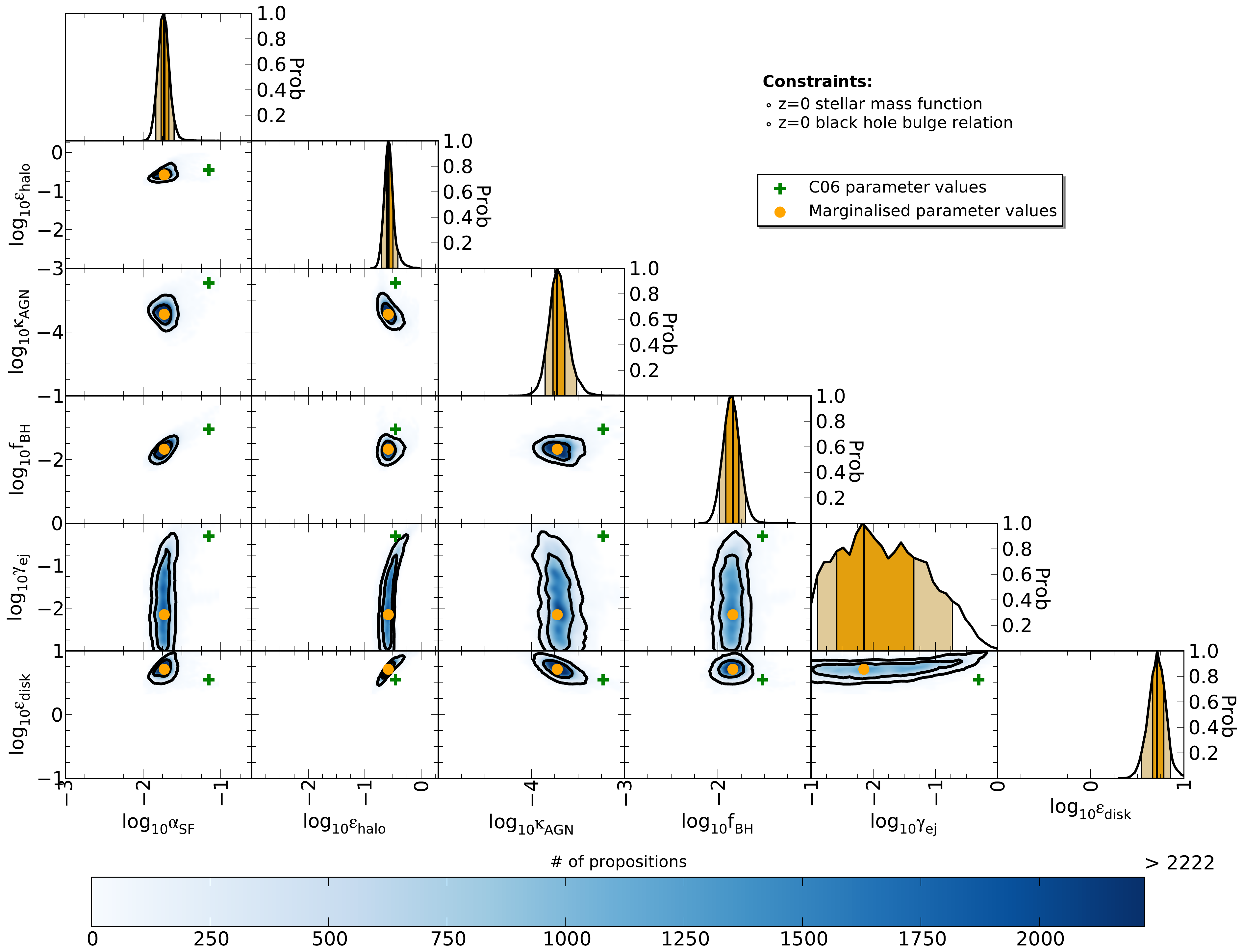}
        \caption{\label{fig:z0_smf-bhbr_prob_matrix} 
            2-D posterior probability distributions (off-diagonal panels) for
            all combinations of the six free model parameters when constraining
            the model against the {\it $z{=}0$ \SMF{} and \BHBR{} simultaneously}.
            Black lines indicate the 1 and 2-$\sigma$ confidence contours.  The
            limits of each panel indicate the prior ranges.  Orange circles show
            the location of the marginalised best values of each
            parameter.  The histograms in the diagonal panels represent the 1-D
            marginalised probability distributions, with the 1 and 2-$\sigma$
            confidence intervals indicated by the dark and light shaded regions
            respectively.  This figure can be directly compared with
          Fig.~\ref{fig:z0_smf_prob_matrix}.}
    \end{minipage}
\end{center}
\end{figure*}

In Fig.~\ref{fig:z0_smf-bhbr_prob_matrix} we show the marginalised posterior
distributions for the six model parameters when constrained against both the
$z{=}0$ \SMF{} and $z{=}0$ \BHBR{} simultaneously.  The joint probability of a particular
model parameter set is determined by calculating the likelihoods for each
constraint individually, as outlined above, and then combining these with equal
weights using standard probability theory:
\begin{equation}
    \label{eqn:L_joint} \mathcal{L}(\theta) = \mathcal{L}_{\mathrm{SMF}}(\theta)
    \cdot \mathcal{L}_{\mathrm{BHBR}}(\theta) 
\end{equation}

As expected, the distributions look similar to those of
Fig.~\ref{fig:z0_smf_prob_matrix}, with the exception that we now also tightly
constrain the values of the merger driven black hole growth and radio mode AGN
feedback efficiencies (\fBH{} and \kappaAGN{}) through the addition of the
information contained in Fig.~\ref{fig:bhbr_prob_panel}.  Unfortunately, we are
still only able to provide an upper limit on the value of the re-incorporation
efficiency parameter, \gammaeject{}.  However, this does tell us that the model
prefers re-incorporation of ejected gas to occur on a timescale longer than
approximately 10 halo dynamical times (roughly equivalent to the Hubble time).

A principal component analysis indicates that the addition of the \BHBR{}
constraint has reduced the number of principal components from two to one.  This
confirms that we have successfully reduced the freedom of the parameter values
with respect to one another.  Again we find that star formation efficiency,
\alphaSF{}, is fully constrained.  However, we now find that supernova cold gas
reheating efficiency, \epsilondisk{}, also contributes practically no
information on the variance of the joint posterior distribution.  This is due to
the fact that we have now restricted the allowed values of the black hole growth
and radio mode AGN heating efficiencies (\fBH{} and \kappaAGN{}), thus
preventing them from compensating for any small shift to \epsilondisk{}, as was
allowed when constraining against the $z{=}0$ \SMF{} alone.

\begin{figure*}
    \begin{center}
    \begin{minipage}{\textwidth}
    \subfigure{\includegraphics[width=0.5\columnwidth]{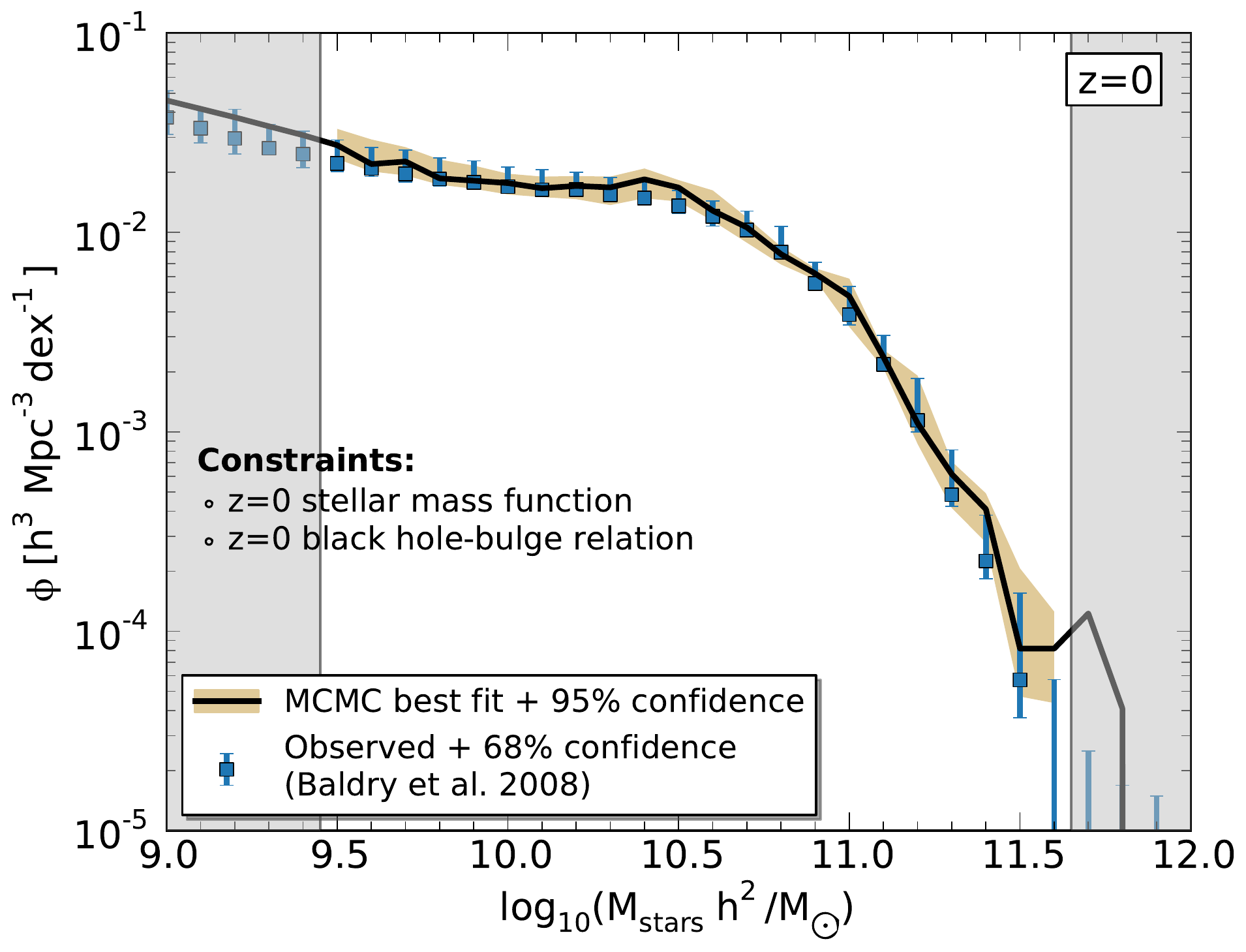}}\quad
    \subfigure{\includegraphics[width=0.5\columnwidth]{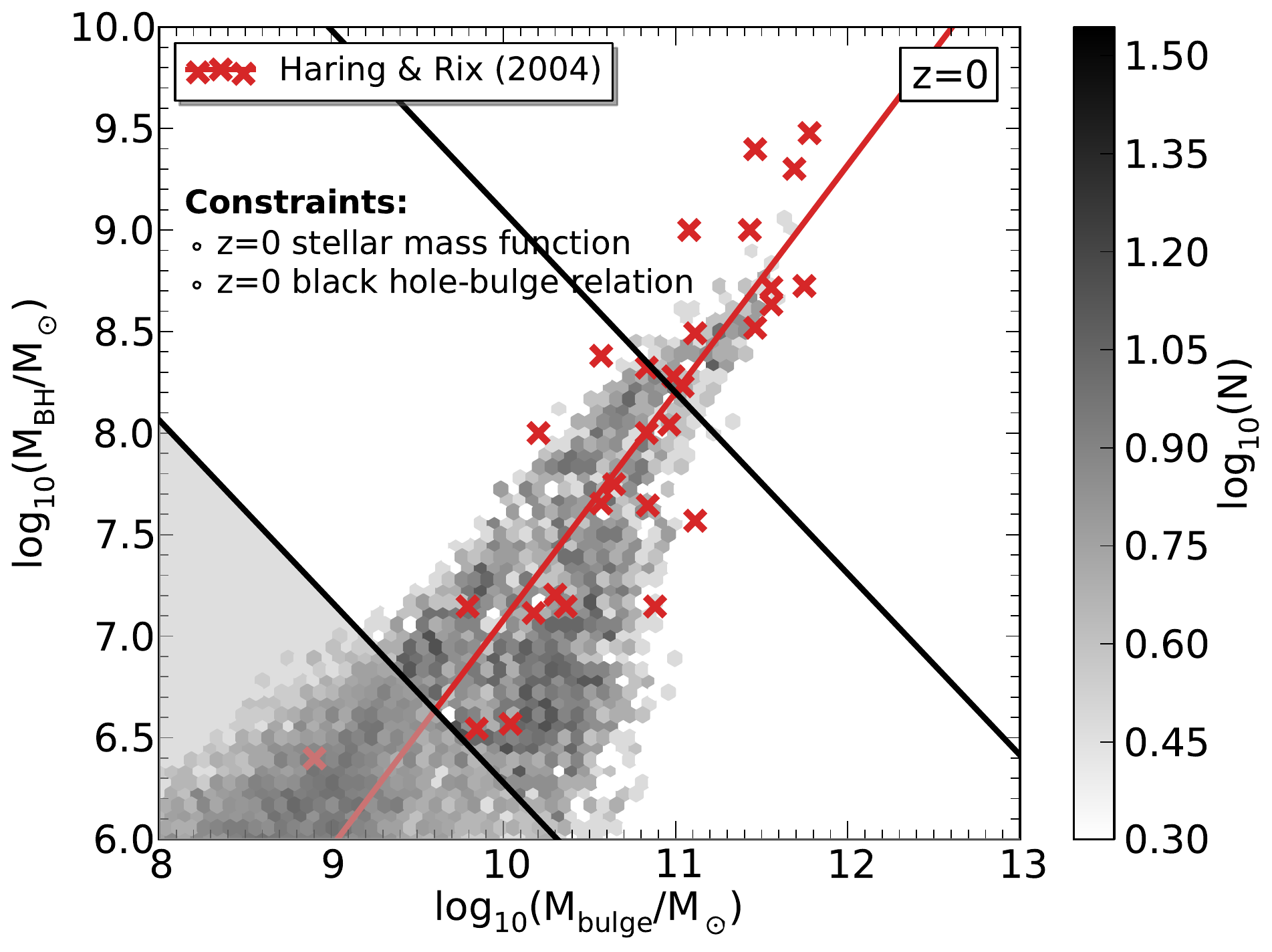}}
    \caption{\label{fig:z0_smf-bhbr_smf} 
      The $z{=}0$ model \SMF{} (left) and \BHBR{} (right) obtained using the
        marginalised best parameter values found when
        constraining against both the {\it $z{=}0$ \SMF{} and \BHBR{}
        simultaneously}.  For the \SMF{}, the solid line with shaded region shows
        the model result along with the 95\% confidence region calculated from the
        posterior distribution.  Blue error bars show the constraining observations
        of \citet{Baldry2008}.  In the right hand panel, the model galaxies are
        indicated by the shaded grey hexbins, while the observational constraint of
        \citet{Haring2004} is shown as the red crosses along with the published best
        fit line.}
\end{minipage}
\end{center}
\end{figure*}

In Fig.~\ref{fig:z0_smf-bhbr_smf} we show the $z{=}0$ \SMF{} and \BHBR{}
obtained using the marginalised best parameters.  These parameter values form
our fiducial $z{=}0$ set and are listed in Table~\ref{tab:params}.

A comparison with Fig.~\ref{fig:z0_smf} indicates that our reproduction of the
observed \SMF{} remains excellent.  However, we note that the likelihood of the
\BHBR{} is only 0.2 when including the \SMF{} constraint.  This is caused by a
slight tension between the preferred parameter values of these two constraining
observations.  A similar result was found by \citet{Henriques2009} when
calibrating their model against the $K$-band luminosity function and \BHBR{}.

Despite this drop in likelihood, statistical agreement within 2$\sigma$ is
still achieved and the resulting \BHBR{} remains cosmetically acceptable.
Finally, we note that a great deal of observational uncertainty remains in the
precise normalisation and slope of the \BHBR{}.  It is therefore possible that
future \BHBR{} measurements will result in a relation that is more easily
reconciled with the \SMF{} in our model.

\subsection{High redshift}
\label{sub:highz_analysis}

\begin{figure}
    \includegraphics[width=\columnwidth]{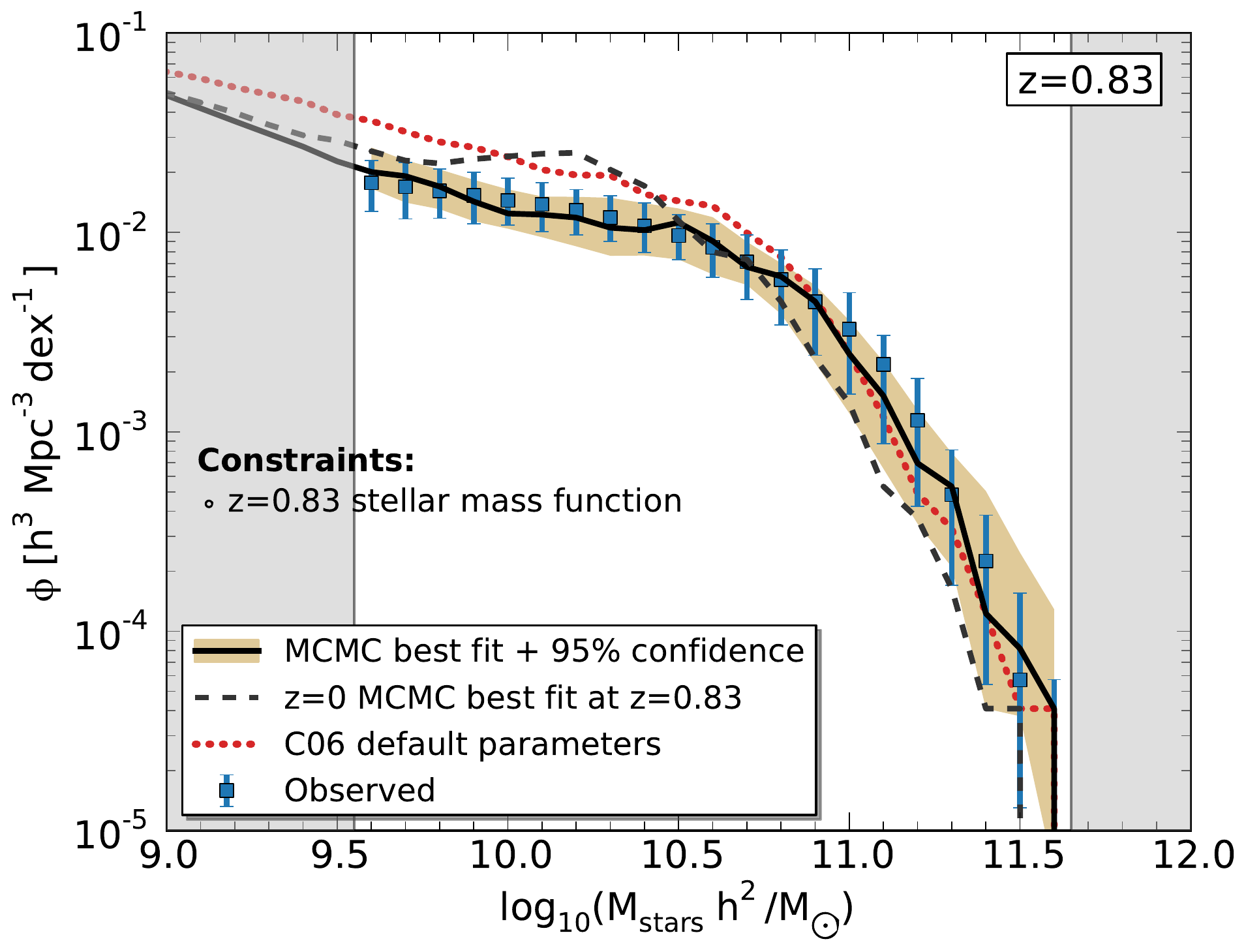} 
    \caption{\label{fig:highz_smf} The resulting $z{=}0.83$ \SMF{} and 95\%
    confidence regions (black line and shaded region) produced by constraining
    the model against the {\it observed $z{=}0.83$ \SMF{}}\/ alone.  The
    constraining observations and 1$\sigma$ uncertainties are shown as blue
    error bars (c.f.~\S\ref{sub:highzSMF}). Also shown for comparison are the
    $z{=}0.83$ stellar mass functions produced by the default \C06{} (red dotted
  line) and $z{=}0$ fiducial (black dashed line) parameter values.}
\end{figure}

In the last section we investigated the constraining power of the $z{=}0$
\SMF{} and \BHBR{}.  These observational quantities allowed us to place strong
restrictions on the values of all but one of the free model parameters.  In this
section we investigate the resulting $z{>}0$ model predictions and also
implement our $z{=}0.83$ \SMF{} constraint on its own
(\S\ref{sub:SMF_constraint}) in order to test the restrictions it imposes on the
model parameters.

In Fig.~\ref{fig:highz_smf} we show the prediction of the default \C06{} model
parameter values (red dotted line), compared against the observed \SMF{} at
$z{=}0.83$ (blue squares; see \S\ref{sub:SMF_constraint} for details).
Similarly to the redshift zero case, the default model over-predicts the number
of galaxies at low masses.  However, it now also predicts a steeper slope to the
high mass end than is observed.

Also shown in Fig.~\ref{fig:highz_smf} is the result obtained using the fiducial
$z{=}0$ parameters of the previous section.  This model predicts an unrealistic
build up of galaxies around $\log_{10}(M\,[h^2/\mathrm{M_{\sun}}]){=}10.25$
which constitutes the population that will later evolve to fill the high mass
end of the distribution at $z{=}0$.  This over-density is a direct result of
under-efficient supernova feedback, allowing lower mass galaxies to hold on to
too much of their cold gas which is subsequently converted into stellar mass.
Overall, the fiducial $z{=}0$ parameters appear to do a worse job of reproducing
the $z{=}0.83$ observations than the original \C06{} values.

\begin{figure}
    \includegraphics[width=\columnwidth]{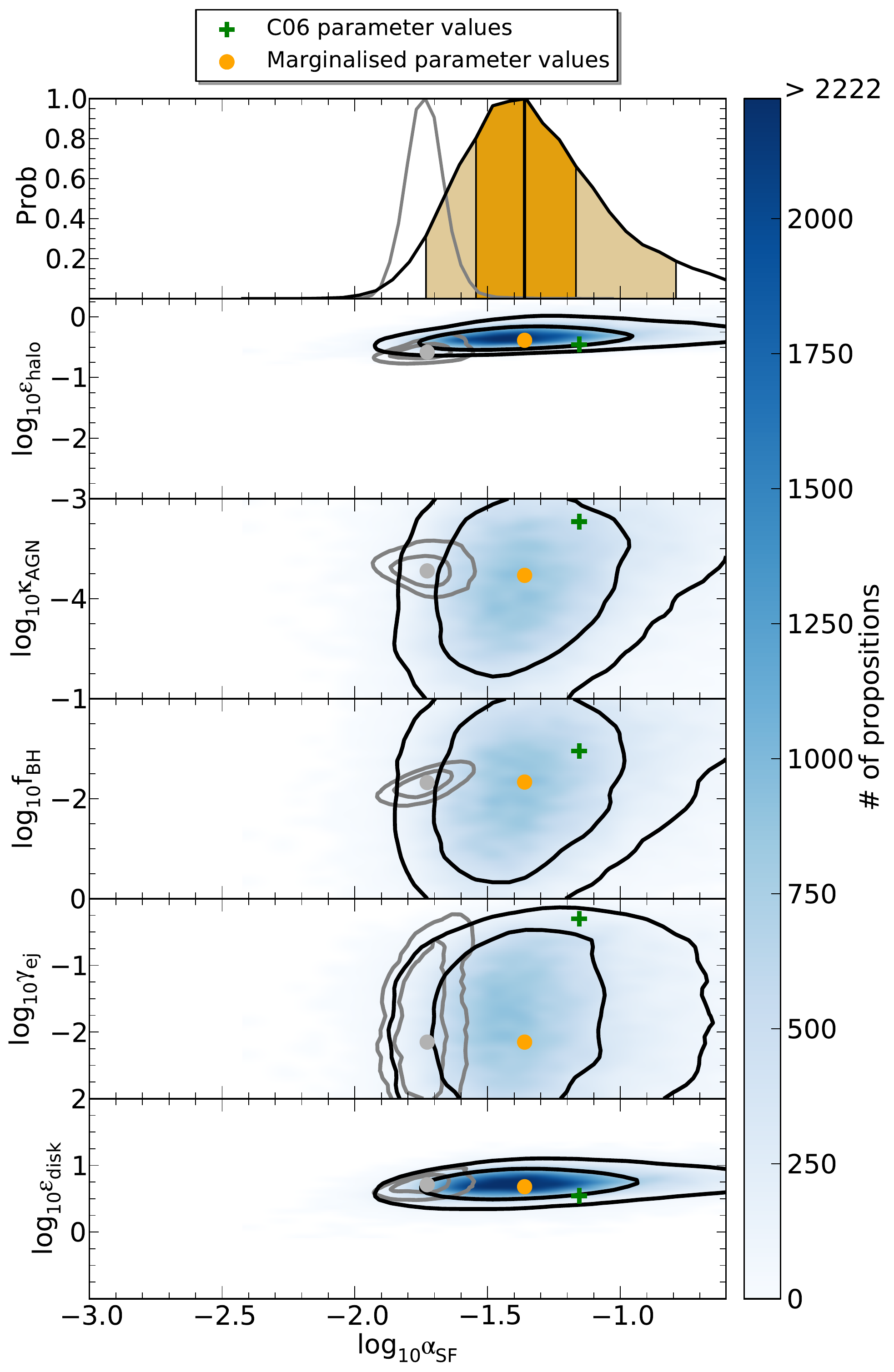}
    \caption{\label{fig:highz_smf_prob_matrix} 2-D posterior probability
    distributions for the star formation efficiency, \alphaSF{}, against the
    five other free model parameters when constraining the model against the
    {\it $z{=}0.83$ \SMF{}}\/ alone.  The limits of each panel are equivalent
    to the MCMC prior ranges.  Orange circles indicate the location of the
    marginalised best values of each parameter.  Black lines represent the 1 and
    2$\sigma$ confidence contours with grey lines indicating the equivalent
    results from Fig.~\ref{fig:z0_smf-bhbr_prob_matrix}.  The histogram in the
    top panel displays the 1-D marginalised probability distribution for
    \alphaSF{}, with the 1 and 2$\sigma$ confidence intervals shown by the dark
    and light shaded regions respectively.}
\end{figure}

The solid black line and shaded region of Fig.~\ref{fig:highz_smf} indicate the
best fit mass function and associated 95\% confidence regions found by
constraining against the observed $z{=}0.83$ \SMF{} alone.  An excellent
agreement can be seen across all masses.  The marginalised best parameter values
and 68\% uncertainties are listed in Table~\ref{tab:params}.  A
subset of the posterior probability distributions are presented in
Fig.~\ref{fig:highz_smf_prob_matrix}.  For comparison, the equivalent 1 and
2$\sigma$ confidence regions of the $z{=}0$ \SMF{}+\BHBR{} constraint from
Fig.~\ref{fig:z0_smf-bhbr_prob_matrix} are also indicated with grey contour
lines.

As was the case when constraining the model against the $z{=}0$ \SMF{} alone
(\S\ref{sub:z0_smf_analysis}), there is little restriction on the values of
black hole growth and radio mode AGN feedback efficiencies (\fBH{} an
\kappaAGN{}).  This is due to the fact that we have no \BHBR{} constraint to
break the degeneracy between these two parameters.  The star formation
efficiency (\alphaSF{}) confidence region is also significantly larger at
$z{=}0$.83 compared to $z{=}0$.  This is primarily a reflection of the
larger observational uncertainties across all stellar masses at this redshift.
Interestingly, we also find that the most likely value of the supernova cold gas
reheating parameter, \epsilondisk{}, shows little evolution between redshifts,
despite the need to increase the upper prior limit on this parameter to account
for a slightly extended probability tail extending past our original upper limit
of \epsilondisk{}=10.

The parameters controlling star formation and supernova halo gas ejection
efficiencies (\alphaSF{} and \epsilonhalo{}) display the largest differences in
posterior distributions with respect to $z{=}0$.  In fact, a principal component
analysis suggests that the only parameter which is truly well constrained by the
$z{=}0.83$ \SMF{} is \alphaSF{}.  Its value is approximately 2.5 times higher
than was the case at $z{=}0$ which is driven by the need to form high mass
galaxies more rapidly in order to achieve a better match to the massive end of
the observed \SMF{}.  However, a side effect is the further build up of galaxies
at intermediate masses ($\log_{10}(M\,[h^2/\mathrm{M_{\sun}}]){=}10.0{-}10.5$)
which must then be alleviated by increasing the strength of the supernova cold
gas ejection efficiency, \epsilonhalo{}.

From Eqn.~\ref{eqn:meject}, we see that the amount of gas ejected from the dark
matter halo entirely by supernova feedback equals zero for $V_{\mathrm{vir}}{>}
V_{\mathrm{vir}}^{\mathrm{cutoff}} {=} V_{\mathrm{SN}}(\epsilon_{\mathrm{halo}} /
\epsilon_{\mathrm{disk}})^{1/2}$.  Hence by increasing \epsilonhalo{} (the halo
hot gas ejection efficiency), we increase the characteristic halo mass at which
supernova feedback becomes ineffective at ejecting gas from the system.  The net
result is a reduction of star formation in more massive galaxies (which
preferentially populate dark matter halos with higher masses, and hence higher
virial velocities) due to a reduced availability of hot gas which can then cool
to fuel star formation.

Given that $V_{\mathrm{vir}}^{\mathrm{cutoff}}$ depends on the ratio of the
supernova ejection and reheating parameters (\epsilonhalo{} and \epsilondisk{}),
why does the $z{=}0$.83 \SMF{} preferentially modify \epsilonhalo{} from its
$z{=}0$ fiducial value instead of \epsilondisk{}?  Again from
Eqn.~\ref{eqn:meject}, we see that a change in \epsilondisk{} results in a
proportional change to the ejected mass.  However, this is independent of the
host halo properties.  On the other hand, modifying \epsilonhalo{} results in a
change to the ejected mass with a magnitude that is inversely proportional to
$V_{\mathrm{vir}}^2$.  Hence increasing \epsilonhalo{} results in both an
increase in the value of $V_{\mathrm{vir}}^{\mathrm{cutoff}}$, as well as
preventing the build-up of excess star forming galaxies just above this halo
velocity where radio mode black hole feedback is still inefficient.

\subsection{Combined redshifts}
\label{sub:multi_z_constraints}

Having presented the results of constraining the model to match observations at
$z{=}0$ and 0.83 individually, we now investigate if it is possible to
achieve a satisfactory result at both redshifts simultaneously.  As shown in
Fig.~\ref{fig:highz_smf_prob_matrix}, there is some tension between the
marginalised posterior distributions at each redshift.  However, it is possible
that a parameter configuration may exist which, although not achieving the best
possible reproduction at either redshift, will still provide a satisfactory
combined result.

\begin{figure*}
    \begin{center}
	\begin{minipage}{0.99\textwidth}
	    \includegraphics[width=0.99\textwidth]{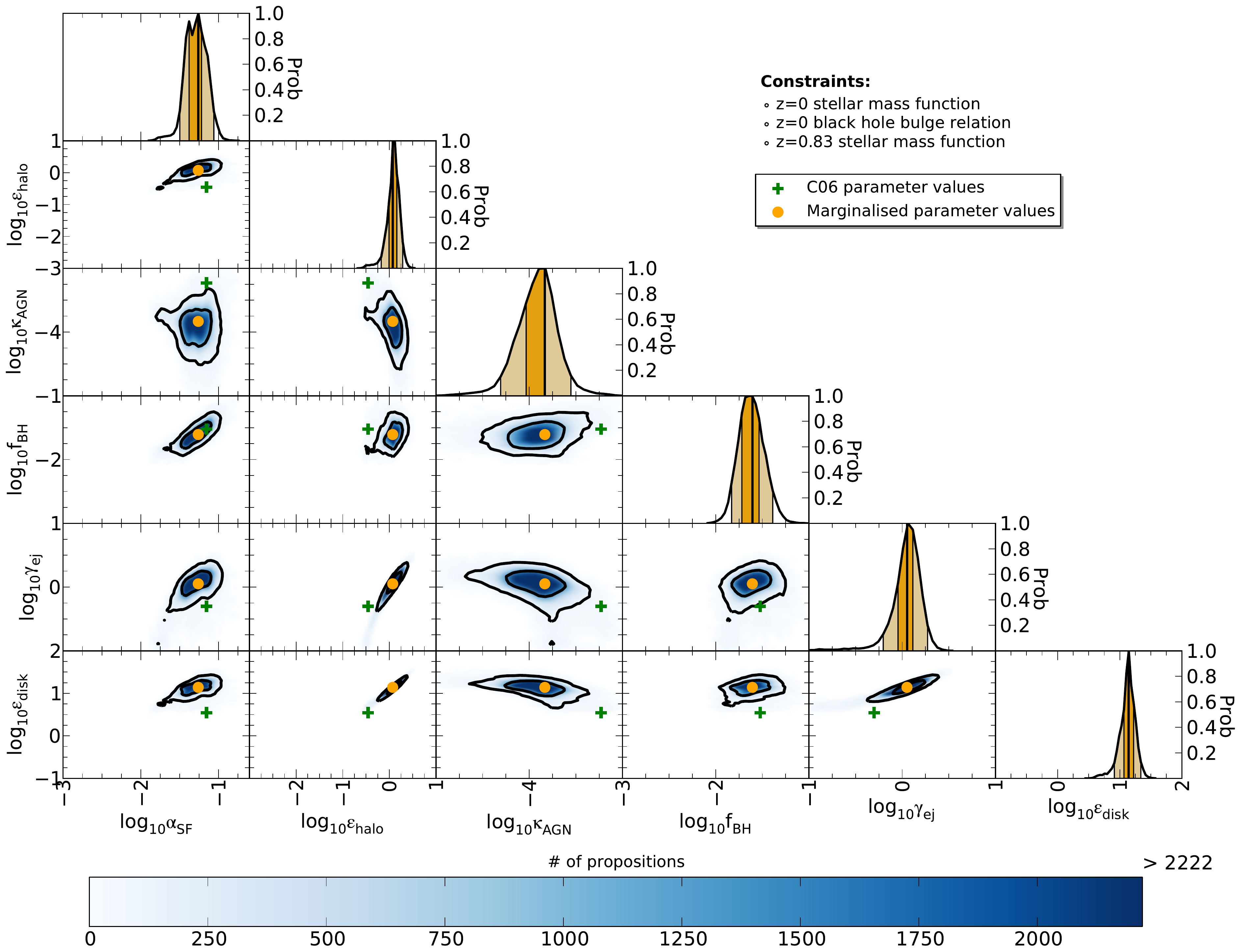}
        \caption{\label{fig:doublez_prob_matrix} 2-D posterior probability
        distributions (off-diagonal panels) for all combinations of the six free
        model parameters when constraining against the {\it $z{=}0$ and
        $z{=}0$.83 \SMF{}s, and the $z{=}0$ \BHBR{}}, all simultaneously.  Black
        lines indicate the 1 and 2-$\sigma$ confidence contours.  The limits of
        each panel indicate our prior ranges.  Orange circles show the location
        of the marginalised best values of each parameter, while red
        diamonds indicate the values from the single parameter set which
        produced the best reproduction of the observations (maximum likelihood
        parameters).  The histograms in the diagonal panels represent the 1-D
        marginalised probability distributions, with the 1 and 2-$\sigma$
        confidence intervals indicated by the dark and light shaded regions
        respectively.  The 1-D maximum likelihood distributions are also shown
        for comparison (red dashed lines).  This figure can be directly compared
      with Figs.~\ref{fig:z0_smf_prob_matrix} and
      \ref{fig:z0_smf-bhbr_prob_matrix}.}
    \end{minipage}
\end{center}
\end{figure*}

Fig.~\ref{fig:doublez_prob_matrix} shows the 1 and 2-D posterior distributions
for the model when constrained simultaneously against the $z{=}0$ \SMF{} and
\BHBR{}, as well as the $z{=}0.83$ \SMF{}.  The 1-D histograms indicate that
this constraint combination places strong restrictions on all of the free model
parameters, including the ejected gas reincorporation rate parameter,
\gammaeject{}. For all previously investigated constraint combinations, the
value of \gammaeject{} has made little impact on the ability of the model to
reproduce the relevant observations.  This has been true for values spanning
several orders of magnitude.  However, when constraining the model to reproduce
two redshifts simultaneously, gas reincorporation rate plays a key role.
 
As discussed in \S\ref{sub:highz_analysis}, the individual redshift constraints
provide strong, but irreconcilable, restrictions on the value of the star
formation efficiency, \alphaSF{}.  When these constraints are combined, the
model is therefore forced to pick one of the preferred \alphaSF{} values and use
the freedom in the other parameters, including \gammaeject{}, to maximise the
joint likelihood.

The main effect of altering the ejected gas reincorporation efficiency,
\gammaeject{}, occurs at the low mass end of the stellar mass function.  It's
only here that supernova feedback is efficient at expelling gas and galaxies have
a significant amount of material in their ejected reservoirs.  In addition, in
our model the ejected mass reservoirs of in-falling satellite galaxies are
immediately incorporated into the hot halo components of their more massive
parents and hence no extra material is added to the ejected component of these
larger systems.  By increasing the value of \gammaeject{} the timescale over
which expelled gas makes its way back into the heating/cooling cycle is
decreased.  This results in more cold gas being available for forming stars in
the lowest mass galaxies, with the net effect being a raising of the low mass
end of the stellar mass function.

As shown in Fig.~\ref{fig:highz_smf}, the fiducial $z{=}0$ parameter set
already produces a \SMF{} at $z{=}0.83$ which is overabundant in low mass
galaxies.  In order for \gammaeject{} to help to alleviate this, its value needs
to be reduced, thus reincorporating less ejected gas into lower mass systems.
Unfortunately however, the marginalised best value of \gammaeject{} using the
$z{=}0$ constraints is already extremely low ($1.7{\times}10^{-3}$) and reducing
it even further has a negligible effect. The model is therefore unable to
utilise this parameter to maximise the joint-redshift likelihood when using the
fiducial $z{=}0$ star formation efficiency.  Fortunately however,
\gammaeject{} can be used to maximise the likelihood achieved with the
$z{=}0.83$ preferred parameters.

Compared to the $z{=}0$ case, the $z{=}0.83$ marginalised best parameters
have a higher star formation efficiency, \alphaSF{}, resulting in a more rapid
transition of galaxies from low to high masses.  As shown in
Fig.~\ref{fig:z0_smf}, the effect on the stellar mass function at $z{=}0$ is
an over-abundance at high stellar masses and a corresponding under-abundance at
low masses.  Increasing the low value of the gas reincorporation rate parameter
(\gammaeject{}) can have a significant effect here by increasing the number of
galaxies below the knee of the mass function.  By also increasing the values of
the supernova feedback gas reheating and ejection parameters (\epsilondisk{} and
\epsilonhalo{}) the model can achieve the correct overall shape whilst moving
the position of the knee by only a small amount.  This allows a better
reproduction of the $z{=}0$ mass function to be achieved.

\begin{figure*}
    \begin{center}
    \begin{minipage}{\textwidth}
    \subfigure{\includegraphics[width=0.5\columnwidth]{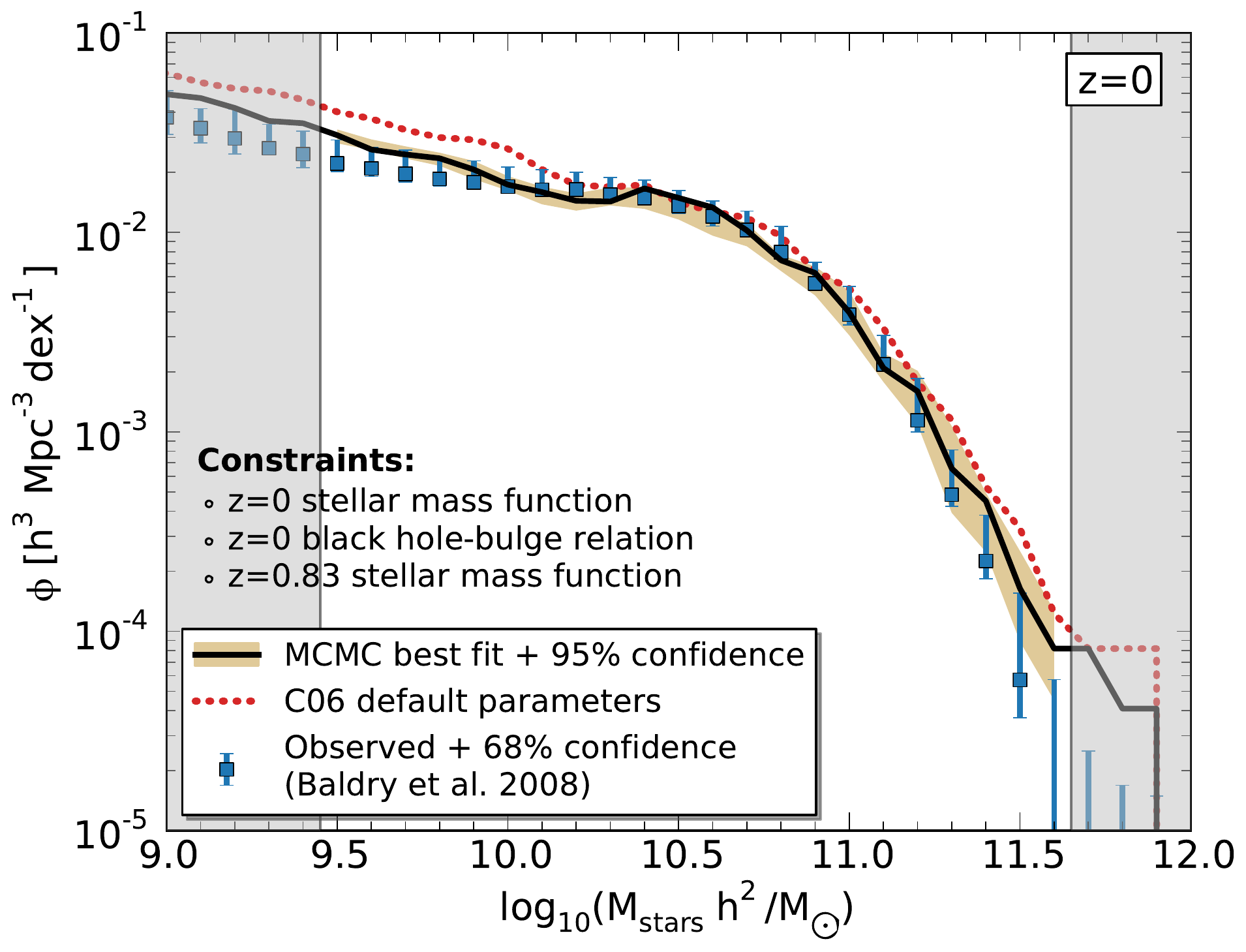}}\quad
    \subfigure{\includegraphics[width=0.5\columnwidth]{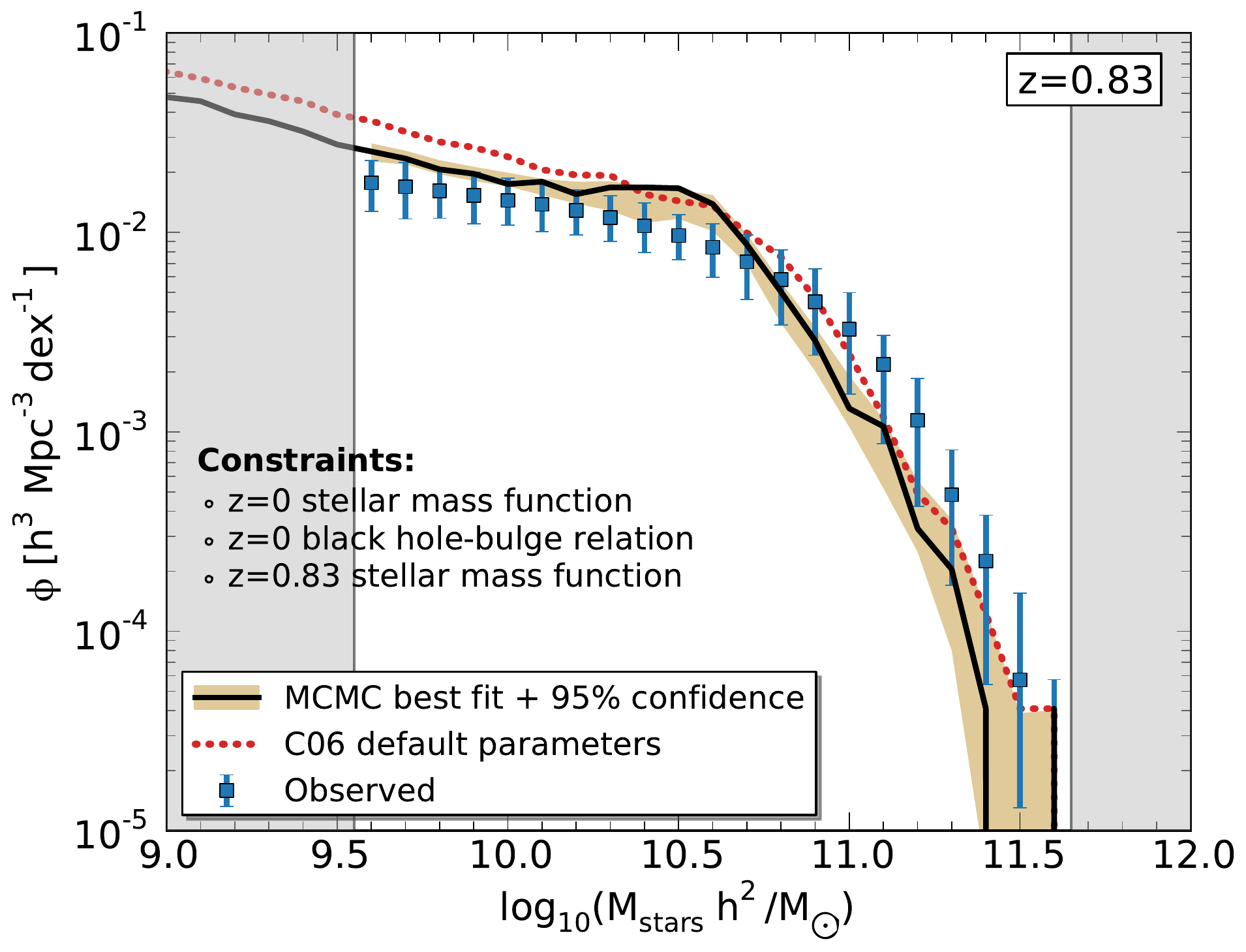}}
    \caption{\label{fig:doublez_smf} The $z{=}0$ (left) and $z{=}0.83$ (right) model
    \SMF{}s obtained using the highest likelihood model parameters when
    constraining against the {\it $z{=}0$ and $z{=}0.83$ \SMF{}s, and the $z{=}0$
    \BHBR{}}, simultaneously.  The solid line with shaded region shows the model
    result along with the 95\% confidence region calculated from the posterior
    distributions.  Blue error bars show the relevant constraining
    observations.}
\end{minipage}
\end{center}
\end{figure*}

The marginalised best values and their uncertainties are presented in
Table~\ref{tab:params}.  Fig.~\ref{fig:doublez_smf} displays the resulting
$z{=}0$ and 0.83 \SMF{}s.  Even though the star formation efficiency parameter
is close to the preferred $z{=}0.83$ value, the changes made to the other
parameters have resulted in a visually poorer reproduction of $z{=}0.83$ mass
function.  However, a reduced $\chi^2$ of 1.28 (with 14 degrees of freedom)
indicates that the fit of the highest likelihood line is still statistically
acceptable.  In order to achieve this level of agreement at both redshifts
simultaneously, we note that we have been required to push the parameters
associated with supernova feedback and reincorporation (\epsilondisk{},
\epsilonhalo{} and \gammaeject{}) to values that are perhaps unrealistic.  We
discuss the interpretation and possible physical implications of this outcome in
the next section.

Finally we note that we have increased the lower limit on the reincorporation
efficiency (\gammaeject{}) prior to 0.1 when applying our joint redshift
constraints.  In testing, we allowed this parameter to go as low as $10^{-3}$,
however, we found that this introduced a large peak in the marginalised
posterior distributions, corresponding to the alternative, but lower
probability, $z{=}0$ preferred star formation efficiency (\alphaSF{}).  As
discussed above, in order to maximise the likelihood achieved using this
\alphaSF{} value, the reincorporation efficiency must be lowered
as much as possible.  However, all values of $\gamma_{\mathrm{ej}} {\la} 0.1$
produce approximately identical joint likelihoods as any sufficiently small
value results in almost no mass of ejected material being reincorporated in to
low mass halos.  When these low mass halos grow sufficiently, they will
eventually reincorporate the material.  However, they will also have set up a
hydro-static hot halo, meaning the effect of recapturing this mass on the
central galaxy will be minimal.

When we allowed the reincorporation efficiency to go to lower values the MCMC
chain spent a large number of successful proposals mapping out the extensive
volume provided by this flat low likelihood feature.  Each of these propositions
contributed to a marginalised peak coinciding with the lower likelihood
$z{=}0$ preferred star formation efficiency.  This feature of the
posterior distribution highlights the importance of selecting suitable priors
which encompass the full range of physically plausible parameter values, but
which do not include large areas of parameter space which are indistinct from
each other due to the details of the model.

If we were to have allowed values of $\gamma_{\mathrm{ej}} {\la}
0.1$ in our presented analysis, we would have unfairly biased our posterior
distributions towards a region of lower likelihood.  It is important to note
however, that given a suitably long chain, the MCMC would eventually still have
converged to provide us with the same posterior distribution peaks as presented
in Fig.~\ref{fig:doublez_prob_matrix}.  Our poor choice of lower prior for
\gammaeject{} would simply have meant that such a convergence would have
required an unfeasibly long chain.

\section{Discussion} 
\label{sec:discussion}

\subsection{Interpreting the joint redshift constraint results}

The fiducial parameter values for our joint redshift constraints (c.f.\@
table~\ref{tab:params}) provide us with a valuable insight into exactly where
the tensions lie within the model when trying to successfully reproduce the late
time growth of stellar mass in the Universe.  The parameters associated with
supernova feedback have been pushed to their limits, and possibly to
unrealistic values.  Using our MCMC analysis of the model when constrained
against each redshift individually allows us to understand the cause of this as
follows:

\begin{itemize}

  \item The values of all of the parameters are driven by the need to put the
    high mass end of the stellar mass function in place as early as possible.
    This is illustrated in Fig.~\ref{fig:highz_smf} where we see that the high
    redshift \SMF{} produced by the $z{=}0$ fiducial parameter values (dashed
    line) under predicts the number density of high mass galaxies.  In order to
    alleviate this discrepancy and provide the best possible match to the
    observations at $z$=0.83 alone, a relatively high star formation efficiency
    is required (see Fig.~\ref{fig:highz_smf_prob_matrix}).

  \item Unfortunately, as shown in Fig.~\ref{fig:z0_smf}, this high star
    formation efficiency causes an under-prediction of the number of low mass
    galaxies at $z{=}0$, due to their rapid growth.  In order to
    counteract this and to provide the best possible result at both redshifts
    simultaneously, the preferred re-incorporation rate of ejected material
    (\gammaeject{}) must be increased. This efficiently increases the number
    density of galaxies with stellar masses below the knee of the mass function
    whilst leaving the high mass distribution unchanged.  Although not
    implausible, it is worth noting that such a high value of \gammaeject{}
    requires the presence of some mechanism to rapidly return ejected material
    into the heating/cooling cycle over timescales close to, or less than,
    the dynamical time of the host dark matter halo.

  \item However, the rise in the number of very low stellar mass galaxies that
    results from such a high gas re-incorporation efficiency is extremely large
    and leads to an overestimation of their number density at both redshifts.
    To compensate, the preferred supernova halo gas ejection
    efficiency (\epsilonhalo{}) is forced to values greater than one, implying
    that either the mean kinetic energy of supernova explosions per unit mass is
    too low, or that some other physical mechanism exists to enhance the
    deposition of this energy into the ISM\@.

  \item Increasing the value of the supernova ejection efficiency to such high
    values allows for the efficient regulation of star formation in larger and
    larger systems, thus pushing the knee of the $z{=}0$ stellar mass function
    to higher masses.  In order to return the knee to its correct position the
    model is forced to equivalently increase the supernova feedback mass loading
    factor (\epsilondisk{}) to ${\approx}14$.  Unfortunately, such a high mass
    loading factor is difficult to reconcile with current observational
    estimates \citep[e.g.][]{Martin1999,Rupke2002,Martin2006} and may suggest
    the need for an additional halo mass dependence for this parameter
    \citep[e.g.][]{Oppenheimer2006,Hopkins2012}.

\end{itemize}

The high values preferred by these parameters suggests that the model
prescriptions for supernova feedback, and possibly gas re-incorporation, are
insufficient.  Similar assertions have been made by other works, most commonly
based upon calibrating semi-analytic models to reproduce the $z{=}0$ stellar
mass or luminosity functions and then investigating the $z{>}0$ predictions
\citep[e.g.][]{Guo2011,Lu2012}.  However, we confirm this finding for the first
time through explicitly attempting to match the observed \SMF{}s at two
redshifts simultaneously.  This allows us to exclude the possibility that a
physically acceptable parameter combination exists within the framework of our
current physical prescriptions.

\subsection{Implications}

In Fig.~\ref{fig:doublez_smf} we show the highest likelihood \SMF{}s produced by
the model when simultaneously constrained against the observed $z{=}0$ and
0.83 \SMF{}s and the $z{=}0$ \BHBR{}.  Although we do manage to achieve
statistically reasonable fits to the \SMF{} at both epochs, there are some clear
tensions at high redshift.  We now discuss the possible implications for our
semi-analytic model as well as for the growth of stellar mass in the Universe.

Despite large observational uncertainties associated with measuring the number
density of the most massive galaxies at $z{\gg}0$, the phenomenon of galaxy
`downsizing', whereby the most massive galaxies in the Universe are in place at
early times, is well established \citep{Heavens2004,Neistein2006}.  Our results
extend those of previous works in suggesting that current galaxy formation
models built upon the hierarchical growth of structure find it difficult to
reproduce the quantitative details of this phenomenon
\citep[e.g.][]{DeLucia2006,Kitzbichler2007,Guo2011,Zehavi2012}.  In particular,
a comparison of the \SMF{}s produced when constraining against each redshift
individually demonstrates that the model struggles to successfully put the
highest mass galaxies in place early on (black dashed line of
Fig.~\ref{fig:highz_smf}) without also under-predicting the number of low mass
galaxies at $z{=}0$ and equivalently over-predicting the number density of the
most massive galaxies (black dashed line of Fig.~\ref{fig:z0_smf}).  

It is possible that the model's under-prediction at high masses may be partially
alleviated by convolving the $z{=}0.83$ model mass function with a normal
distribution of a suitable width in order to account for systematic
uncertainties in the observed stellar masses \citep{Kitzbichler2007,Guo2011}.
We do not carry out such a procedure here, as at least some fraction of this
uncertainty should be included in our constraining observations and we do not
wish to add a further add-hoc correction without justification.  However, it may
prove impossible for the model to self-consistently reproduce the observed
stellar mass function at multiple redshifts without including these additional
observational uncertainties \citep{Moster2012}.

In our model, star formation, supernova feedback and gas re-incorporation are
assumed to proceed with a constant efficiency as a function of halo mass across
the full age of the Universe.  However, our findings could be interpreted as
suggesting the need to incorporate an explicit time dependence to these
efficiencies; in particular to provide a preferential increase to the rate of
stellar mass growth in massive halos in the early Universe.  This would help
establish the high mass end of the stellar mass function early on without
over-producing the number of lower mass galaxies at late times.

Unfortunately, adding further layers of parametrisation to current processes,
such as an explicit time dependence, makes the interpretation of model results
increasingly difficult.  This is especially so when attempting to uncover the
relative importance of physical processes that shape the evolution of different
galaxy populations.  To combat this, it is important to ensure that all new
additions to a semi-analytic model have a strong physical motivation.  

Having said that, modifications to the rate of growth of stellar mass in the
early Universe is supported by other studies.  In particular, it has been
suggested that the star formation efficiency of galaxies must peak at earlier
times for more massive galaxies \citep[e.g.][]{Moster2012}.  This could possibly
represent a number of physical processes such as a metallicity dependent star
formation efficiency \citep{Krumholz2012} or a rapid phase of stellar mass
growth due to high redshift cold flows \citep{Dekel2009}.  Alternatively, the
apparent need to incorporate an explicit time dependence to the star formation
and feedback efficiencies could signify an overestimation of the merger
timescales in the model at early times \citep{Weinmann2011}.  Also, we note that
star formation proceeds in our model following a simple gas surface density
threshold argument.  It may be the case that we are over predicting the size of
the most massive galaxies at early redshift and therefore under-predicting the
level of star formation in these objects.  We do not specifically track the
build up of angular momentum in our simulated galaxies, instead relying on the
spin of the parent dark matter halo as being a good proxy.  It is unlikely that
this assumption is valid at high redshift
\citep[e.g.][]{Dutton2009,Kimm2011,Brook2011} and, even if it were, the low time
resolution of our input simulation coupled with the low number of particles in
halos at these times might mean that the halo spin values are systematically
unreliable here.

Rather than suggesting the need for a time dependent star formation efficiency,
an additional interpretation of our findings could be the need to include an
intra-cluster light component (ICL)  in the model
\citep{Gallagher1972,Purcell2007,Guo2011}.  Sub-halo abundance matching studies
have suggested that mergers involving massive galaxies may result in significant
fractions (${\ga}80\%$) of the in-falling satellite mass being deposited in to
this diffuse component rather than ending up in the central galaxy
\citep{Conroy2007,Behroozi2012}.  Since the majority of late time growth of the
most massive galaxies is heavily dominated by mergers, the inclusion of such a
stellar mass reservoir may allow the effective suppression of the growth of
these massive objects, reducing the need to push the supernova feedback
parameters to such extreme values.

Finally, the majority of modern galaxy formation models are now able to
reproduce many of the most important observational quantities of the $z{=}0$
Universe.  Any changes to the underlying cosmology or input merger tree
construction can typically be accounted for by varying the free model
parameters.  However, achieving the correct time evolution of the full galaxy
population is a more difficult task and makes us far more dependant on the
details of dark matter structure growth.  If this growth does not correctly
match the real Universe then this is something that the models will try to
counteract, possibly leading us to conclusions about the baryonic physics which
could be incorrect.  To fully assess the level to which missing or poorly
understood baryonic physics are responsible for discrepancies from observed
galaxy evolution, a detailed analysis of the effects of cosmology, dark matter
halo finding, and merger tree construction on the output of a single galaxy
formation model is needed.

\section{Conclusions}
\label{sec:conclusions}

In this work we investigate the ability of the \C06{} semi-analytic galaxy
formation model to reproduce the late time evolution of the growth of galaxies
from $z{\approx}0.8$ to the present day.  In particular we focus on matching
the $z{=}0$ and $z{=}0.83$ \SMF{}s as well as the $z{=}0$ \BHBR{}.  To
achieve this we utilise \MCMC{} techniques, allowing us to both statistically
calibrate the model against the relevant observations and to investigate the
degeneracies and tensions between different free parameters.

Our main results can be summarised as follows:
\begin{itemize}
    \item The \C06{} model is able to provide a good agreement with the \SMF{}
      and \BHBR{} at $z{=}0$ (\S\ref{sub:z0_analysis};
      Figs.~\ref{fig:z0_smf_prob_matrix},~\ref{fig:z0_smf}).  However, when
      attempting to match both simultaneously there are some minor tensions
      between the favoured parameter values (\S\ref{sub:z0_analysis};
      Figs.~\ref{fig:bhbr_prob_panel},~\ref{fig:z0_smf-bhbr_smf}).
    \item The model is also able to independently provide a good agreement with
      the observed \SMF{} at $z{=}0$.83.  In order to achieve this, a higher
      star formation efficiency is necessary than was preferred to match the
      $z{=}0$ constraints.  This is to ensure that the massive end of the
      \SMF{} is entirely in place by this early time, as required by our
      constraining observations (\S\ref{sub:highz_analysis};
      Figs.~\ref{fig:highz_smf},~\ref{fig:highz_smf_prob_matrix}).
    \item When attempting to reproduce the observations at both redshifts
      simultaneously, a number of tensions in the model's physical prescriptions
      are highlighted.  In particular, the struggle to reconcile the high star
      formation efficiency required to reproduce the high mass end of the \SMF{}
      at $z{=}0.83$ with the observed increase in normalisation of the low
      mass end at $z{=}0$ (\S\ref{sub:multi_z_constraints};
      Fig.~\ref{fig:doublez_smf}).
    \item Our attempts to model the evolution of galaxies at $z{\la}0.8$
      suggest that the supernova feedback prescriptions of the model may be
      incomplete, possibly requiring the addition of extra processes that
      preferentially enhance star formation in the most massive galaxies at
      $z{>}1$ (\S\ref{sec:discussion}).
\end{itemize}

This is the first time that a full semi-analytic model, based on the input of
N-body dark matter merger trees, has been statistically calibrated to try and
reproduce a small focussed set of observations at multiple redshifts
simultaneously.  Only by carrying out this procedure, and fully exploring the
available parameter space of our particular model, are we able to conclusively
demonstrate that the model struggles to match the late time growth of the galaxy
stellar mass function.  Our analysis also provides us with important insights as
to what changes may be necessary to alleviate these tensions.  Having said that,
despite requiring somewhat unlikely parameter values, we do achieve a
statistically reasonable fit to the observations at both redshifts
simultaneously.  For the purpose of producing mock galaxy catalogues at
$z\la0.8$, this best fit model is perfectly adequate.

For this work, we have only considered the \SMF{} and \BHBR{}.  However, it is
likely that the addition of extra observational constraints will further help to
isolate the parts of the model which require particular attention.  For example,
the gas mass--metallicity relation \citep{Tremonti2004} is particularly
sensitive to the re-incorporation efficiency due to its ability to regulate the
dilution of a galaxy's gas component with low metallicity material ejected at
early times.  In future work we will extend our analysis to redshifts greater
than one and investigate the constraints provided by other quantities
such as the mass--metallicity relation as well as the baryonic Tully-Fisher
relation, galaxy colour distribution and stellar mass density evolution.

Finally, we also stress that our results and conclusions are sensitive to the
magnitude of the uncertainties associated with our observational constraints.
Although we have endeavoured to ensure their accuracy, it is likely that they
may still be underestimated.  For example, the high mass end of $z{\gg}0$
stellar mass functions are heavily susceptible to cosmic variance effects due to
the deep observations required to simultaneously resolve galaxies at lower mass
scales (typically done in smaller fields).  Some works have also suggested
systematic uncertainties of 0.3 dex or more when estimating stellar masses via
broad-band photometry \citep{Conroy2009}.  Furthermore, the
magnitude of many of these uncertainties increases significantly with redshift,
and can result in errors of up to 0.8 dex around the knee of the measured
stellar mass functions at $z=1.3{-}2$ \citep{Marchesini2009}.  More detailed
comparisons between state-of-the-art semi-analytic models and high redshift
observations will therefore require us to greatly improve our measurements at
these redshifts.

\section*{Acknowledgements}

SM acknowledges the support of a Swinburne University SUPRA postgraduate
scholarship.  Both SM and GP are supported by the ARC Laureate Fellowship
of S. Wyithe.  GP also acknowledges support from the ARC DP programme.  DC
acknowledges receipt of a QEII Fellowship awarded by the Australian government.

The authors would like to thank E. Taylor and K. Glazebrook for a number of
helpful discussions. The Millennium Simulation used as input for
the semi-analytic model was carried out by the Virgo Supercomputing Consortium
at the Computing Centre of the Max-Planck Society in Garching. Semi-analytic
galaxy catalogues from the simulation are publicly available at
http://www.mpa-garching.mpg.de/millennium/.

\bibliographystyle{mn2e}
\bibliography{master}

\label{lastpage}

\end{document}